\documentclass{IEEEtran}
\IEEEoverridecommandlockouts

\input{Format.tex}
\DeclarePairedDelimiter\parens{\lparen}{\rparen}
\DeclarePairedDelimiter\bracks{\lbrack}{\rbrack}
\DeclarePairedDelimiter\braces{\lbrace}{\rbrace}
\DeclarePairedDelimiter\abs{\lvert}{\rvert}
\DeclarePairedDelimiter\norm{\lVert}{\rVert}

\DeclarePairedDelimiter\ceil{\lceil}{\rceil}

\DeclareMathOperator{\diag}{\textnormal{diag}}

\newcommand*{\trn}{\!^{\mathsf{T}}}
\newcommand*{\hmt}{\!^{\mathsf{H}}}
\newcommand*{\inv}{\!^{-1}}

\DeclareMathOperator{\C}{\mathbb{C}}

\newcommand*{\E}[1]{\mathbb{E}\bracks*{#1}}

\newcommand*{\set}[1]{\braces*{\,#1\,}}

\makeatletter
\DeclareRobustCommand\onedot{\futurelet\@let@token\@onedot}
\def\@onedot{\ifx\@let@token.\else.\null\fi\xspace}

\def\eg{\emph{e.g}\onedot} 
\def\ie{\emph{i.e}\onedot}

\makeatother

\DeclareMathOperator{\sinc}{\textnormal{sinc}}

\newtcolorbox{stretchbox}[1][]{
    height fill,
    colback=white,
    colframe=black,
    #1
    }

\newtcolorbox{problem}[1]{
    breakable,
    toptitle=2.5pt,
    bottomtitle=2.5pt,
    colbacktitle=white,
    coltitle=black,
    colback=white,
    colframe=black,
    fonttitle=\large\bfseries,
    title={#1 \hfill Grade:\hspace*{0.15\paperwidth}\ }, 
}

\newtcolorbox{solution}[1]{
    breakable,
    colback=black!3,
    fonttitle=\bfseries,
    title={#1},
}

\newcommand\redout{\bgroup\markoverwith{\textcolor{red}{\rule[.5ex]{2pt}{0.4pt}}}\ULon}

\allowdisplaybreaks
\usetikzlibrary{shapes,arrows,positioning,calc,angles,quotes,patterns.meta,decorations.pathreplacing}
\RequirePackage{comment}

\title{\textcolor{black}{Beyond the Delay-Doppler Domain: A Time-Frequency Framework for Low-Overhead,  Scalable OTFS Channel Estimation}
\thanks{The work was supported by National Science Foundation (NSF) grant ECCS-2514270 and Army Research Office (ARO) grant \textcolor{black}{W911NF261A271}.}
}
\author{
Kailong~Wang~\orcidlink{0000-0002-3415-0790},~\IEEEmembership{Student~Member,~IEEE},
Athina~Petropulu~\orcidlink{0000-0001-7380-7815},~\IEEEmembership{Fellow,~IEEE}%
\thanks{The authors are with the Department of Electrical and Computer Engineering, Rutgers University, Piscataway,
NJ, 08854 USA (e-mail: kailong.wang@rutgers.edu; athinap@rutgers.edu).}%
}

\newcommand{\mPhi}{\boldsymbol{\Phi}}

\newcommand{\mI}{\mathbf{I}}

\newtheorem{theorem}{Theorem}

\newtheorem{proposition}{Proposition}
\newtheorem{lemma}{Lemma}

\begin{document}
\maketitle

\begin{abstract}
\textcolor{black}{
Most works on pilot-aided orthogonal time frequency space (OTFS) channel estimation operate in the delay-Doppler (DD) domain, where embedded-pilot guard regions must be sized to the unknown maximum delay-Doppler spread, resulting in high overhead that is compounded by antenna scaling: per-antenna guard regions grow overhead with the number of transmit antennas, while shared guard regions avoid this but introduce pilot contamination. Fractional Doppler further reduces DD-domain sparsity, increasing estimation complexity. This paper shows that, unlike the DD representation, the time-frequency (TF) representation provides a low-overhead, scalable framework for OTFS channel estimation based on the physical scatterer parameters (angle, delay, Doppler, and gain). We derive exact TF- and DD-domain input–output relationships under fractional Doppler. The TF expression decomposes the received signal into a desired component, identical in form for both integer and fractional Doppler, and an explicitly characterized interference term; this Doppler-invariant structure, absent from the DD representation, enables scalable TF-domain estimation. This motivates a TF-domain pilot design based on private TF bins protected by TF/DD guard bins, which preserves cross-antenna pilot orthogonality while incurring overhead that depends only on the number of pilots, not the array size. Building on this, we develop a low-complexity coarse-to-fine estimation method combining discrete Fourier transform (DFT)-based coarse estimation with dimensionality-reduced sparse recovery. The exact DD-domain expression, in contrast, provides the signal model needed for data-symbol recovery and explains why the complexity of DD-domain estimation grows substantially under fractional Doppler. Simulations show the proposed method achieves accurate channel state information (CSI) estimation under both integer and fractional Doppler, remains robust to pilot pollution, and scales to large arrays with substantially lower overhead and complexity than existing schemes.}
\end{abstract}

\begin{IEEEkeywords}
Integrated Sensing and Communication,  OTFS, Pilot-Aided Channel Estimation, Interference Analysis
\end{IEEEkeywords}

\section{Introduction}

\IEEEPARstart{T}{he} proliferation of high-mobility communication scenarios, such as vehicle-to-everything (V2X) and unmanned aerial vehicle (UAV) networks, has made achieving reliable, high-capacity, and low-latency wireless transmission a primary goal for 5G and beyond networks. A key challenge in these environments is the rapid time variation of wireless channels~\cite{saad20206g}. Conventional orthogonal frequency-division multiplexing (OFDM) signaling~\cite{Weinstein2009ofdm,tao2025securitydirectionalmodulationtime} degrades under such conditions due to the resulting Doppler effects, which cause inter-carrier interference (ICI).
To mitigate these issues, the Orthogonal Time Frequency Space modulation (OTFS) scheme~\cite{hadani2017OrthogonalTime, raviteja2018InterferenceCancellation} has emerged as a promising technology due to its inherent robustness to Doppler shifts and its proven superior performance in fast-fading channels.
Unlike conventional OFDM, which maps data to the time-frequency (TF) domain, OTFS distinguishes itself by multiplexing symbols in the delay-Doppler (DD) domain. 
In this domain, the channel impulse response of a highly mobile environment is transformed into a sparse, quasi-time-invariant representation characterized by a few key DD parameters. 
This enables accurate equalization and signal detection in high-mobility scenarios.

\textcolor{black}{The primary challenges in the OTFS channel estimation literature are the high estimation overhead, which typically increases with the number of antennas, and fractional Doppler, which arises whenever a channel scatterer's true Doppler shift does not align with the DD grid, a condition that is exacerbated under the short frame durations required for low-latency operation.
Under fractional Doppler, the received signal leaks along the Doppler dimension, with most of its power concentrated around the nearest integer Doppler index. Early Single-Input Single-Output (SISO) OTFS channel estimation schemes embed a single DD-domain pilot within a zero guard region to prevent pilot–data interference and recover the sparse DD-channel response via threshold-based detection~\cite{raviteja2019EmbeddedPilotAided}. 
However, under fractional Doppler, reliably detecting the leaked components in neighboring Doppler bins requires excessively high pilot power, \eg, $50$~dB~\cite{raviteja2019EmbeddedPilotAided}, which is often impractical.
\cite{wei2022OffGridChannel, tang2024NovelOffGrid, guo2025LowComplexity, sun2025LowComplexity, li2025FractionalDelayDoppler} employ multiple uncorrelated pilots along with enhanced detection algorithms, reducing the pilot-power requirement.
Later works~\cite{lee2025SingleToneBasedChannel, wang2025ModelDrivenChannel} reduce the requirement back to a single low-power pilot, albeit at the cost of increased algorithmic complexity.
As a departure from DD-domain pilots,~\cite{zhang2024lowoverheadotfs} reformulates the OTFS system as a precoded OFDM system, enabling DD-domain channel estimation via guarded frequency-domain or time-domain embedded pilots that estimate a dispersive frequency-Doppler channel matrix through interpolation.
In all the aforementioned methods, the required guard region introduces substantial signaling overhead and reduces the achievable communication rate. 
Moreover, because the guard region must cover the maximum delay–Doppler spread, which is generally unknown a priori, it is often conservatively oversized, further increasing the overhead.}

\textcolor{black}{
Extending SISO embedded-pilot schemes with guard regions to Multiple-Input Multiple-Output (MIMO) systems is challenging because preserving pilot orthogonality across transmit antennas generally requires antenna-specific guard regions, scaling the pilot overhead with array size. Accordingly,~\cite{kollengoderamachandran2018MIMOOTFSHighDoppler} extends the aforementioned works by employing non-overlapping guard regions per antenna, but the resulting signaling overhead increases proportionally with the number of transmit antennas, limiting scalability.}
MIMO methods utilizing multiple embedded DD-domain pilots with overlapping guard regions across antennas have also been proposed~\cite{shen2019ChannelEstimation, wang2025AdaptiveBlock, zhao2022BlockSparse, liang2024TwoDimensionalDelayDoppler}, 
\textcolor{black}{
which, again, recover each path from a small set of neighboring DD bins around its dominant integer-grid location.
In those methods, the pilots are independent across antennas, ensuring they remain separable at the receiver even when overlapping. However, these methods retain the guard-region overhead limitations of SISO-embedded pilot schemes, since the guard region size still depends on the unknown maximum delay and Doppler spread.}
More critically, the overlapped guard region design suffers from pilot pollution, \ie, interference among overlapping DD pilots across antennas, degrading performance as the number of transmit antennas increases~\cite{shen2019ChannelEstimation, liang2024TwoDimensionalDelayDoppler}. 

%

There are also studies that eliminate the need for guard regions by superimposing pilots on data symbols, using either single~\cite{yuan2021dataaidedChannel, mishra2022OTFSChannel, sheikh2025BayesianLearningaided} or full-grid~\cite{liu2022BEMOTFS,huang2024MPBased} superimposed DD-domain pilots for SISO systems, with MIMO extensions presented in~\cite{mehrotra2023dataaidedCSI}. These approaches employ computationally intensive successive interference cancellation (SIC), \ie, iteratively estimating the channel and refining it with coarsely recovered symbols until the data symbol's effect is eliminated from the received pilot, thereby increasing latency.
\textcolor{black}{Despite the development of numerous low-overhead detection methods~\cite{raviteja2018LowcomplexityIterative, yuan2020SimpleVariational, singh2022LowComplexityLMMSE, xu2025LowComplexity}, the SIC required by superimposed-pilot schemes may limit their scalability to practical MIMO-OTFS systems.}

\textcolor{black}{
In summary, {practical and scalable OTFS channel estimation {based on} the DD response}  remains challenging because embedded pilots incur substantial guard-region overhead, overlapped pilots suffer from pilot–data interference, and joint channel-and-data recovery incurs high computational complexity. 
\textcolor{black}{
We argue that this is rooted in the DD-domain channel model, which converts the Doppler-induced phase variation in the time or TF domain into a shift along the Doppler dimension. 
{Although this yields a time-invariant channel model under high Doppler, it comes at the cost of} representing the received DD-domain signal as a superposition of multiple shifted copies of the transmitted symbol grid, with each copy corresponding to a different propagation path. As a result, the pilots and data symbols become heavily intertwined, especially in the presence of fractional Doppler.}
This motivates the exploration of channel-estimation methods in other domains.
Existing studies have investigated time-domain methods in which pilots are inserted between, rather than embedded within, OTFS bursts.
Both SISO channels~\cite{srivastava2021BayesianLearning}, and MIMO channels~\cite{srivastava2022DelayDopplerAngular, srivastava2022BayesianLearning, mehrotra2024onlinebayesian} have been considered.
While this design avoids pilot–data interference, it does not provide intra-burst channel state information (CSI), which is essential in the high-mobility scenarios targeted by OTFS.
However, to the best of our knowledge, no TF-domain channel parameter estimation method for OTFS has been developed with mathematical support, largely because the high Doppler-induced ICI has been considered difficult to characterize in the TF domain.
}

\textcolor{black}{
Independent of the estimation domain, OTFS channel estimation methods can also be classified according to the quantities they estimate.
One class estimates the full CSI matrix directly, while another models the wireless channel as a collection of discrete scatterers and formulates channel estimation as the recovery of their angle-of-arrival (AoA), angle-of-departure (AoD), delays, Doppler shifts, and complex gains.
The latter class, referred to here as  \emph{parametric}, assumes that the channel consists of a small number of significant propagation paths, a valid assumption in high-frequency communications~\cite{saad20206g, hadani2017OrthogonalTime, raviteja2018InterferenceCancellation}. 
Compared with the former class~\cite{raviteja2019EmbeddedPilotAided}-\cite{mehrotra2023dataaidedCSI}, which approximates each channel path using a small support centered on the nearest integer Doppler bin, introducing a small but unavoidable loss of channel information, the parametric approach exploits the channel's inherent sparsity, requiring substantially fewer pilots than estimating the CSI matrix entry independently.
It further enables estimation of delay and Doppler beyond the resolution of the DD grid~\cite{zegrar2024OTFSBasedISAC,zegrar2025FractionalDelay}, 
and yields physically interpretable quantities directly useful for sensing applications.}

\textcolor{black}{
In this paper, we propose an embedded-pilot-aided, low-overhead, scalable, parametric channel estimation framework for MIMO-OTFS systems applicable to both integer and fractional Doppler. Unlike existing methods, which primarily rely on approximate DD-domain models, we derive new exact TF- and DD-domain input–output (I/O) expressions that directly involve the channel's angle, delay, Doppler, and gain parameters, and show that the TF domain provides a more suitable framework for channel estimation. The derived TF expression reveals that the desired response has the same structure under both integer and fractional Doppler, unlike its DD-domain counterpart. Although each TF observation is contaminated by interference from other TF symbols, our analysis shows that this interference can be explicitly characterized and effectively controlled, so that it does not prevent the channel parameters from being accurately estimated from the TF observations. This result challenges the common belief that high-Doppler ICI is intractable in the TF domain and provides a rigorous mathematical foundation for TF-domain channel estimation, enabling a low-overhead, scalable TF-domain pilot design.
The proposed pilot placement builds on private TF bins~\cite{wang2025ISACMIMO}, eliminating the need for large guard regions and thereby reducing pilot overhead, which depends solely on the number of pilots rather than on the number of transmit antennas. It also preserves pilot orthogonality across transmit antennas, thus avoiding inter-antenna pilot contamination, without compromising the invertibility of the TF–DD mapping. In particular, an explicit rank condition ensures that the selected DD guard-bin locations preserve invertibility.
Building on this design, we develop a coarse-to-fine channel estimation framework that combines discrete Fourier
transform (DFT)-based coarse parameter estimation, pilot-based virtual-array construction, and sparse signal recovery (SSR). The coarse estimates narrow the parameter space and substantially reduce the size of the SSR dictionary matrix, enabling high-resolution, low-complexity channel parameter estimation even for large-scale systems. We also derive the exact DD-domain I/O, which provides the signal model required for symbol recovery, and also explains why DD-domain channel estimation complexity increases substantially under fractional Doppler.}

\color{black}
The contributions of this paper are summarized as follows.

\noindent C1) We derive exact TF- and DD-domain I/O relationships for OTFS systems with practical rectangular pulse shaping, matched filtering, and fractional Doppler. 
The TF-domain expression resembles the ideal bi-orthogonal I/O by showing that each propagation path’s response reduces to a pure complex exponential across the subsymbol index, scaled by a subsymbol-independent complex coefficient whose form is identical for integer and fractional Doppler, with the remaining signal energy captured by an explicitly characterized additive interference term. 
It also establishes the TF domain as a Doppler-invariant framework for scalable channel parameter estimation.
The DD-domain expression, in contrast, provides the exact signal model required for symbol recovery and further explains why the complexity of DD-domain channel estimation increases substantially under fractional Doppler.

\noindent C2) Motivated by the exact TF I/O, we propose a scalable TF pilot design based on private TF bins, protected by TF guard bins, and by DD guard bins whose placement is guaranteed to preserve invertibility via a rank condition. The proposed pilot placement preserves pilot orthogonality across transmit antennas, is robust to pilot contamination, and incurs a signaling overhead that depends only on the number of pilot symbols, independent of the number of transmit antennas. Furthermore, we show that the interference observed on private pilot bins can be modeled as additive zero-mean noise and derive the corresponding signal-to-interference-plus-noise ratio (SINR), demonstrating that reliable estimation is achieved with only a modest increase in pilot power.

\noindent C3) Using the proposed TF pilots, we develop a coarse-to-fine channel estimation framework with substantially reduced complexity. 
Exploiting the subsymbol-independent coefficients established by the TF I/O, we obtain coarse AoA estimates via a DFT across receive antennas, and coarse Doppler and delay estimates by arranging pilots into time and frequency arms to enable low-complexity DFT-based processing. 
The coarse AoA and delay estimates are then used to infer coarse AoD estimates via bistatic geometric constraints.
We exploit the orthogonality of the proposed TF pilots to construct a virtual array and formulate an SSR problem to refine the channel parameter estimates. We develop a dimensionality-reduction framework that transforms the 4D SSR problem into a sequence of much smaller SSRs by combining the coarse estimates, bistatic geometric constraints, and the separable structure of the steering vectors.
 

\color{black}

\smallskip
\noindent\textit{Relation to the literature.} 

The proposed CSI estimation method is inspired by our recent work ~\cite{wang2025ISACMIMO, wang2025BandwidthEfficient, wang2025VirtualArray, wang2025VirtualArrays}, where sensing with a colocated MIMO receiver was considered, and the concept of TF private bins was introduced to construct a virtual array. In~\cite{wang2025ISACMIMO, wang2025BandwidthEfficient, wang2025VirtualArray, wang2025VirtualArrays}, target estimation can be viewed as sensing-channel estimation, implemented by a colocated receiver based on knowledge of the transmitted symbols. In this paper, CSI estimation can similarly be viewed as sensing-channel estimation, but conducted by a well-separated communication receiver and implemented based on knowledge of the transmitted pilots rather than the transmitted symbols.
%
\textcolor{black}{This paper uses the concept of TF private bins as a building block to develop a novel channel estimation scheme, enabled by a new exact expression for the TF-domain signal. This scheme would not have been possible based on the approximate TF expression of \cite{wang2025ISACMIMO}. Furthermore, the invertible TF/DD pilot-and-guard construction, interference analysis, data-recovery formulation, and receiver-side coarse-to-fine CSI estimator together constitute a new analytical and algorithmic framework.}

A preliminary study was reported in~\cite{wang2025LowOverhead},
where we assumed that the transmitter’s parameter estimates could be conveyed to the receiver. 
This implies that the receiver relies on potentially outdated CSI for equalization. 
In fast-varying channels, accurate receiver-side channel estimation is essential.
\textcolor{black}{Unlike transmitter-side estimation with a known waveform, receiver-side CSI acquisition relies on limited pilots, requiring efficient pilot design.}
Furthermore, unlike~\cite{wang2025LowOverhead}, this work considers rectangular transmit pulse shaping and receiver matched filtering, while also accounting for fractional Doppler.


\smallskip

\noindent\textit{Paper organization.}
The remainder of this paper is organized as follows.
\textcolor{black}{
In \cref{sec:MIMO_OTFS}, we review the SISO-OTFS system model and modulation/demodulation process.
In \cref{sec:TF_MIMO_analysis}, we derive exact TF- and DD-domain I/O for practical OTFS systems.
In \cref{sec:pilot}, we introduce the proposed TF-domain pilot design based on private TF bins, and analyze the resulting pilot SINR.}
In \cref{sec:coarse_estimates}, we present a low-complexity algorithm for obtaining coarse estimates of the channel parameters from the proposed TF pilots.
In \cref{sec:SSR}, we describe the virtual-array construction and develop a coarse-to-fine SSR framework for refining the channel parameter estimates.
In \cref{sec:modified_sfft}, the recovery of transmitted symbols is presented.
In \cref{sec:overhead}, the pilot-overhead and computational complexity of the proposed framework are analyzed and compared with those of existing methods.
Simulation results are presented in \cref{sec:simulation}, and concluding remarks are provided in \cref{sec:conclusion}.

\section{OTFS System Background}\label{sec:MIMO_OTFS}

This section provides background on a SISO wireless system transmitting OTFS waveforms.
The transmitter sends packet bursts of duration $T=N\Delta t$ and bandwidth $B=M\Delta f$, where $N$ is the number of subsymbols, $M$ is the number of subcarriers, $\Delta t$ is the subsymbol duration, and $\Delta f$ is the subcarrier spacing.
The orthogonality condition requires that $\Delta t \cdot \Delta f = 1$~\cite{raviteja2018InterferenceCancellation}.
In each burst, a set of $NM$ symbols is arranged on the DD grid,
\begin{align*}
    \set{\bracks*{k\Delta \nu, l\Delta\tau} \mid k=0,1,\ldots,N-1;l=0,1,\ldots,M-1},
\end{align*}
where $k$ and $l$ are Doppler and delay indices.
The grid spacing is \(\Delta \nu={1}/{(N\Delta t)}\) and \(\Delta \tau ={1}/{(M\Delta f)}\).

The DD channel model consisting of $J$ scatterers is~\cite{raviteja2018InterferenceCancellation}
\begin{align}\label{eq:SISO_channel}
    {h}(\tau,\nu) 
    &= \sum_{j=0}^{J-1}\beta_j\delta(\tau-\tau_j)\delta(\nu-\nu_j),
\end{align}
where $\beta_j$ is the complex channel attenuation factor, $\tau_j$ is the delay corresponding to range $R_j$ ($\tau_j=R_j/c$), and $\nu_j$ is the Doppler corresponding to radial speed $v_j$ sensed by the receiver ($\nu_j=v_j/\lambda$) of the $j$-th scatterer, respectively.
Different paths are mutually independent with power $\sigma_\beta^2 = \mathbb{E}\left[|\beta_j|^2\right]$.
Each scatterer's Doppler and delay do not necessarily align with the DD grid, and 
the $j$-th scatterer's Doppler and delay are 
expressed as
$\nu_{j}=(k_{j}+\kappa_j)\Delta\nu$, $\tau_{j}=l_{j}\Delta\tau$,
where $k_j$, $l_j$ are integers and $\kappa_j\in[-1/2,1/2]$ denotes fractional Doppler. 
In wideband systems, the impact of fractional delays can be neglected~\cite{raviteja2018InterferenceCancellation} since the delay spacing $\Delta\tau$ is typically sufficiently small; however, the impact of fractional Doppler is significant since the Doppler spacing $\Delta\nu$ is large for low latency operation.
The dimensions of the grid satisfy $N\geq 2\max\abs{k_j}+1$ and $M\geq\max l_j$ to support all Doppler shifts and delays present.
The DD channel is sparse~\cite{hadani2017OrthogonalTime, raviteja2018InterferenceCancellation};
therefore $J$ is small.

Let $x[k,l]$ be the symbol placed on DD bin $[k,l]$.
The symbols are mapped to the TF domain via the inverse symplectic finite Fourier transform (ISFFT)~\cite{raviteja2018InterferenceCancellation},
\begin{align}\label{eq:ISFFT}
    X[n,m] 
    &= \frac{1}{\sqrt{NM}}\sum_{k=0}^{N-1}\sum_{l=0}^{M-1}x[k,l]e^{i2\pi\bracks*{\frac{kn}{N}-\frac{ml}{M}}}.
\end{align}
The analog signal for transmission, $s(t)$, is created via the Heisenberg transform~\cite{raviteja2018InterferenceCancellation}, \ie,
\begin{align}\label{eq:Heisenberg}
    s(t) 
    &\!=\! \sum_{n=0}^{N-1}\sum_{m=0}^{M-1}X[n,m]g_{tx}(t-n\Delta t)e^{i2\pi m\Delta f(t-n\Delta t)},
\end{align}
where $g_{tx}(t)$ is the rectangular pulse function with amplitude $1/\sqrt{\Delta t}$ for $t\in[0,\Delta t]$ and $0$ at all other times.
Before transmission, a cyclic prefix (CP) of length greater than  $\max l_j$ is appended to $s(t)$ to avoid interference between OTFS blocks.

The signal that reaches the receiver can be expressed as
\begin{align}\label{eq:rx}
    r(t)
    &=\sum_{j=0}^{J-1} s(t-\tau_j) \beta_j e^{i2\pi\nu_j(t-\tau_j)}+w_t(t),
\end{align}
where $w_t(t)$ is the time-domain additive white Gaussian noise (AWGN) with power \textcolor{black}{$\sigma_{w}^2$}.
After filtering with the matched filter $g_{cx}(t)=g_{tx}(t)$, and sampling over a duration $T$ at frequency $B$ (\ie, applying the Wigner transform~\cite{raviteja2018InterferenceCancellation}), the signal is mapped to the TF domain with expression given in the next section.

\section{Exact OTFS I/O Expressions}\label{sec:TF_MIMO_analysis}

\color{black}


This section derives exact TF- and DD-domain input--output relationships for practical OTFS systems. 
The TF-domain expression establishes a Doppler-invariant framework for scalable channel estimation, while the DD-domain expression provides the exact signal model required for symbol recovery. 


\color{black}

\subsection{Exact Expression of Channel I/O in the TF Domain}

\begin{theorem}\label{thm:Y}
The TF I/O expression for the OTFS system is
\begin{align}\label{eq:signal}
    \!Y[n,m]
    \!=\! {X[n,m]} \!\sum_{j=0}^{J-1}\! \beta_j 
    H^j[n,m] 
    \xi_j
    \!+\!I[n,m]\!+\!W[n,m] 
\end{align}
where 
\begin{align}
    &H^j[n,m]
    \!=\! e^{-i2\pi\frac{(k_j+\kappa_j)l_j}{NM}}e^{i2\pi\bracks*{\frac{(k_j +\kappa_j)n }{N} - \frac{ml_j}{M}}}, \!&&\! \label{eq:TFChannel} \\
    &\xi_j
    = \textcolor{black}{
    \begin{cases}
    \frac{M-l_j}{M}, & k_j+\kappa_j=0, \\
    \frac{1}{M} \frac{e^{i2\pi\frac{(k_j+\kappa_j)l_j}{NM}}-e^{i2\pi\frac{(k_j+\kappa_j)}{N}}}{1-e^{i2\pi\frac{(k_j+\kappa_j)}{NM}}}, \label{eq:xi_rest} & \text{otherwise,}
    \end{cases}
    }\\
    &I[n,m] = \sum_{j=0}^{J-1} \beta_j e^{-i2\pi\frac{(k_j+\kappa_j)l_j}{NM}} e^{i2\pi\frac{(k_j+\kappa_j)n}{N}} 
    \times \label{eq:TFinterference}\\
    &\frac{1}{M} \biggl\{\sum_{\substack{m'=0\\ m'\neq m}}^{M-1} X[n,m'] e^{-i2\pi\frac{m'l_j}{M}} \sum_{l=l_j}^{M-1} 
    e^{i2\pi\bracks*{\frac{(k_j+\kappa_j)l}{NM}-\frac{(m-m')l}{M}}}
    \notag \\
    &+ \sum_{m'=0}^{M-1} X[n-1,m'] e^{-i2\pi\frac{m'l_j}{M}} \sum_{l=0}^{l_j-1} 
    e^{i2\pi\bracks*{\frac{(k_j+\kappa_j)l}{NM}-\frac{(m-m')l}{M}}}
    \biggr\}, \notag
\end{align}
and $W[n,m]$ is  TF-domain AWGN. 
\end{theorem}
The proof of \cref{thm:Y} is detailed in \cref{sec:Appendix_Y}.

\color{black}

\textcolor{black}{
\Cref{eq:signal} provides a unified treatment of integer- and fractional-Doppler channels, since the desired response, $H^j[n,m]\xi_j$, has the same functional form in both cases.
}

Since the DD-domain symbols are independent, the unitary nature of the SFFT guarantees that the TF symbols are uncorrelated.
Consequently, the additive terms in \cref{eq:signal} are mutually uncorrelated, and the power of the TF symbol splits exactly among the three terms.

\begin{proposition}\label{prop:power}
The power of interference is the exact complement of the desired signal power $\sigma_\beta^2\sum_{j=0}^{J-1}\abs*{\xi_j}^2$, \ie,
\begin{equation}
    \E{\abs*{I[n,m]}^2} = \sigma_\beta^2\sum_{j=0}^{J-1}(1-\abs*{\xi_j}^2).
\end{equation}
\end{proposition}

The proof is given in \cref{sec:appendix_I_SISO}.

\Cref{thm:Y,prop:power} provide several important insights into the TF-domain representation.

\textit{Insight 1 (Physical interpretation of $\xi_j$):}
{The coefficient $\xi_j$ has a clear physical interpretation as the complex gain of the desired TF-domain response of the $j$-th propagation path.
\(|\xi_j|^2\) quantifies the fraction of the path energy preserved in the desired TF bin, while the complementary fraction $1-|\xi_j|^2$ leaks into the interference term $I[n,m]$ (for proof see \cref{eq:includedterm,eq:excludedterm,sec:appendix_I_SISO}).} 
To interpret \(|\xi_j|^2\), define the normalized delay and Doppler
\begin{align}
    \rho_j \triangleq \frac{l_j}{M}, \qquad
    \varrho_j \triangleq \frac{{k_j+\kappa_j}}{N}.
\end{align}
When $k_j+\kappa_j=0$ (static scatterer), it holds that
\[
|\xi_j|^2=(1-\rho_j)^2.
\]
For a dynamic scatterer, \ie, $k_j+\kappa_j\neq0$, it holds that
\begin{align}
    |\xi_j|^2
    =
    \left(
    \frac{\sin\!\left(\pi(1-\rho_j)\varrho_j\right)}
    {M\sin\!\left(\pi\varrho_j/M\right)}
    \right)^2. \label{xi}
\end{align}
For wideband systems, \ie, for large $M$, \cref{xi} becomes
\begin{align}
|\xi_j|^2
\approx
(1-\rho_j)^2
\sinc^2\!\left((1-\rho_j)\varrho_j\right).
\end{align}
where we used the approximation 
\(M\sin\!\left(\frac{\pi\varrho_j}{M}\right) \approx \pi\varrho_j \)

Under practical assumptions,  the maximum normalized delay satisfies $\rho_j\leq0.07$, consistent with the normal-CP duration specified in 3GPP~\cite{ETSI_TS138211_V1530_2018}, and the normalized Doppler analogously satisfies $\abs*{\varrho_j}\leq 0.07$, $\abs*{\xi_j}^2 \gtrsim 0.85$, indicating that at least $85\%$ of each path's power remains in the desired signal.
Note that $\rho_j=\tau_j\Delta f, \ \varrho_j=\frac{v_jf_c}{c\Delta f}$ do not depend on the specific grid used, but rather on the physical properties of the channel.
The aforementioned bounds $\rho_j\leq0.07$ and $\abs*{\varrho_j}\leq0.07$ cover two realistic, standardized 5G deployment regimes: an FR1 macro-cell channel ($f_c=4$~GHz, $\Delta f=30$~kHz, $316$~ns root-mean-square (RMS) delay spread) and an FR2 indoor-factory channel ($f_c=28$~GHz, $\Delta f=120$~kHz, $150$~ns RMS delay spread), yielding $\rho\approx 0.028$ and $\rho\approx 0.054$, respectively\footnote{{A channel's actual maximum delay spread is often cited as $3\times$ the RMS delay spread for outdoor macro-cell models \cite{ETSI_TS138211_V1530_2018}.}}, with corresponding mobility ceilings of $157.5$~m/s for FR1 and $90$~m/s for FR2.


\textcolor{black}{\textit{Insight 2 (Subsymbol-Index-Independent $\xi_j$):} Unlike the approximate TF model 
\color{black} in~\cite{wang2025ISACMIMO}, where the coefficient $\xi_j[n]$ depends on the subsymbol index, the exact derivation yields a coefficient $\xi_j$ that is independent of $n$. Consequently,  the response of each propagation path remains a pure complex exponential across the subsymbol index. This property enables low-complexity DFT-based parameter estimation in \cref{sec:coarse_estimates}.}

\textit{Insight 3 (Equivalent ideal-pulse representation):}
\Cref{thm:Y} shows that the TF-domain I/O under rectangular pulse shaping can be expressed as the ideal bi-orthogonal TF input--output relationship~\cite[Eq.~(13)]{raviteja2018InterferenceCancellation}, scaled by $\xi_j$ and accompanied by an additive interference term $I[n,m]$. This means that the  desired signal preserves the ideal TF structure, ensuring that the physical channel parameters $\tau_j$ and $\nu_j$ remain identifiable.




\color{black}

%


\textcolor{black}{
The following theorem provides the exact DD I/O expression for symbol detection (\cref{sec:modified_sfft}) after channel estimation.
}

\begin{theorem}\label{thm:y}
The received DD-domain signal is 
\begin{align}\label{eq:DDIO_rect}
    y[k,l]
    &=
    \sum_{j=0}^{J-1} 
    \sum_{\substack{q=k_j-\\ (N-1)}}^{k_j} x[[k-(k_j-q)]_N,[l-l_j]_M] \notag
    \\
    &\quad \times
    \beta_je^{-i2\pi\frac{(k_j+\kappa_j) l_j}{NM}} \frac{1}{N} \alpha_j[k,l,q] +w[k,l], 
\end{align} 
where $[\cdot]_N$ denotes the modulo $N$ operation and
\begin{align}\label{eq:phase_DD}
\!&\!
\alpha_j[k,l,q] = \notag \\
\!&\!
\begin{cases}
    \gamma_j(q)e^{i2\pi\frac{(k_j+\kappa_j) l}{NM}}, \!&\! l_j \!\leq\! l \!<\! M, \\
    (\gamma_j(q)-1)e^{i2\pi\frac{(k_j+\kappa_j) l}{NM}}e^{-i2\pi\frac{[k-(k_j-q)]_N}{N}}, \!&\! 0 \!\leq\! l \!<\! l_j,
\end{cases} \\
\!&\!
\gamma_j(q)\stackrel{\triangle}{=}\sum_{n=0}^{N-1}e^{-i2\pi(-q-\kappa_j)\frac{n}{N}},
\end{align}
and $w[k,l]$ is DD-domain AWGN.
\end{theorem}
The proof of \cref{thm:y} is provided in \cref{sec:appendix_y}.


\cref{thm:y} provides the exact DD-domain I/O for practical OTFS systems. 
It shows that the complexity of DD-domain processing depends critically on Doppler grid alignment. 
For integer-Doppler channels, \cref{eq:DDIO_rect} reduces to the well-known simplified DD-domain model by eliminating the summation over $q$~\cite[Sec.~IV-B]{raviteja2018InterferenceCancellation}. 
In contrast, fractional Doppler channels preserve this summation, which represents Doppler-axis power leakage and substantially increases the complexity of DD-domain channel estimation. This observation contrasts with \cref{thm:Y}, whose TF-domain representation maintains the same structure for both integer and fractional Doppler. 
Consequently, the TF domain provides a unified framework for channel estimation, while the exact DD-domain expression remains useful for 
recovering the transmitted symbols.


\color{black}

Fractional delay is often neglected in wideband systems~\cite{raviteja2018InterferenceCancellation}.
To see why, 5G NR FR1 ($B\approx100$~MHz) and FR2 ($B\approx400$~MHz)
provide one-way range resolutions of approximately $3$~m and $0.8$~m,
respectively, allowing path delays to be approximately aligned with the
integer delay grid after synchronization.
Nevertheless, we show that the TF-domain I/O remains advantageous over the
DD-domain I/O in the presence of fractional delay.

\begin{lemma}\label{lemma:frac_delay}
The OTFS TF-domain I/O under fractional delay, i.e.,
$\tau_j=(l_j+\varepsilon_j)/(M\Delta f)$, is
\begin{multline}
    Y'[n,m]
    =
    \sum_{j=0}^{J-1}
    \beta_j
    e^{-i2\pi\frac{(k_j+\kappa_j)(l_j+\varepsilon_j)}{NM}}
    e^{i2\pi\bracks*{
        \frac{(k_j+\kappa_j)n}{N}
        -
        \frac{m(l_j+\varepsilon_j)}{M}
    }} \\
    \xi_j
    X[n,m]
    + I[n,m] + W[n,m],
\end{multline}
where
\begin{align}
    \xi_j
    =&
    \begin{cases}
    \displaystyle
    \frac{M-\ceil*{l_j+\varepsilon_j}}{M},
    & k_j+\kappa_j=0,
    \\
    \frac{1}{M}
    \frac{
        e^{i2\pi\frac{(k_j+\kappa_j)\ceil*{l_j+\varepsilon_j}}{NM}}
        -
        e^{i2\pi\frac{k_j+\kappa_j}{N}}
    }{
        1-e^{i2\pi\frac{k_j+\kappa_j}{NM}}
    },
    & \text{otherwise},
    \end{cases}
    \label{eq:xi_frac_delay}
    \\
    I[n,m]
    =&
    \sum_{j=0}^{J-1}
    \beta_j
    e^{-i2\pi\frac{(k_j+\kappa_j)(l_j+\varepsilon_j)}{NM}}
    e^{i2\pi\frac{(k_j+\kappa_j)n}{N}}\\
    \frac{1}{M}
    \biggl\{
    &\sum_{\substack{m'=0\\m'\neq m}}^{M-1}
    X[n,m']
    e^{-i2\pi\frac{m'(l_j+\varepsilon_j)}{M}} \notag\\
    &\sum_{l=\ceil*{l_j+\varepsilon_j}}^{M-1}
    e^{i2\pi\bracks*{
        \frac{(k_j+\kappa_j)l}{NM}
        -
        \frac{(m-m')l}{M}
    }}
    \notag\\
    +
    &\sum_{m'=0}^{M-1}
    X[n-1,m']
    e^{-i2\pi\frac{m'(l_j+\varepsilon_j)}{M}} \notag\\
    &\sum_{l=0}^{\ceil*{l_j+\varepsilon_j}-1}
    e^{i2\pi\bracks*{
        \frac{(k_j+\kappa_j)l}{NM}
        -
        \frac{(m-m')l}{M}
    }}
    \biggr\}.
    \label{eq:I_frac_delay}
\end{align}
Here, $\ceil*{\cdot}$ denotes the ceiling function.
\end{lemma}

The OTFS TF-domain I/O expression under fractional delay in
\Cref{lemma:frac_delay} can be derived straightforwardly along the lines of
\Cref{thm:Y}.
Specifically, the first sample belonging to the current subsymbol is indexed by
$\ceil*{l_j+\varepsilon_j}$, which determines the summation limits in
\eqref{eq:xi_frac_delay} and \eqref{eq:I_frac_delay}, whereas the unquantized
delay $l_j+\varepsilon_j$ remains in the TF-domain phase.
Setting $\varepsilon_j=0$ yields
$\ceil*{l_j+\varepsilon_j}=l_j$, thereby recovering the integer-delay
expression in \Cref{thm:Y} exactly.
Consequently, the physical interpretation of $\xi_j$ remains unchanged:
it characterizes the attenuation of the desired TF-domain component caused by
the loss of orthogonality due to the channel delay and Doppler.

Conversely, fractional delay introduces additional power leakage along the
delay dimension in the DD-domain I/O, as also observed in
\cite{wei2022OffGridChannel,li2025FractionalDelayDoppler}, yielding a denser
effective channel matrix and increasing the difficulty of nonparametric
DD-domain CSI estimation.
In the following, we therefore consider only fractional Doppler, since
fractional delay does not qualitatively alter the structure of the TF-domain I/O.

\color{black}

\subsection{Extension to MIMO-OTFS Systems}
Let us now extend the above results to the case of a MIMO system comprising a transmitter with $N_t$ antennas (TX) and a receiver with $N_c$ antennas (RX). 
The carrier frequency is $f_c\,\unit{\Hz}$, and the wavelength is $\lambda=c/f_c$ with $c$ being the speed of light.
The transmit and receive antennas form uniform linear arrays (ULAs) with spacing $g_t$ and $g_c$, respectively.

We employ the angular DD-domain channel model~\cite{srivastava2022DelayDopplerAngular}, consisting of $J$ scatterers at AoA $\theta_j$ and AoD $\varphi_j$,
\ie, 
\begin{align}\label{eq:channel_model}
    \mathbf{h}(\theta,\varphi,\tau,\nu) 
    \!=\! \sum_{j=0}^{J-1}\beta_j\mathbf{a}_{c}(\theta_j)\mathbf{a}_{t}\hmt(\varphi_j)\delta(\tau-\tau_j)\delta(\nu-\nu_j),
\end{align}
where $\mathbf{a}_{c}(\theta_j)$ is the AoA steering vector and $\mathbf{a}_{t}(\varphi_j)$ is the AoD steering vector. Other setups are the same as \cref{sec:MIMO_OTFS}.

The received TF symbol on bin $[n,m]$ at the $n_c$ antenna is
\begin{multline}\label{eq:TFIO_rect_desired}
    Y_{n_c}[n,m]
    = \sum_{j=0}^{J-1} 
    e^{-i2\pi n_cg_c \frac{\sin\theta_j}{\lambda}} 
    \sum_{n_t=0}^{N_t-1} {X_{n_t}[n,m]} 
    \\
    \times e^{-i2\pi n_tg_t \frac{\sin\varphi_j}{\lambda}} \beta_j e^{-i2\pi\frac{(k_j+\kappa_j) l_j}{NM}} e^{i2\pi\bracks*{\frac{(k_j +\kappa_j)n }{N} - \frac{ml_j}{M}}} \xi_j  \\
    + I_{n_c}[n,m] + W_{n_c}[n,m].
\end{multline}
The first term of \cref{eq:TFIO_rect_desired} extends \cref{eq:signal} by incorporating transmit and receive spatial steering vectors for each multipath component.
The term $I_{n_c}[n,m]$ extends \cref{eq:TFinterference} by incorporating transmit and receive spatial steering vectors for each multipath component.
In the following, the interference term $I_{n_c}[n,m]$ and the additive noise are neglected for analytical tractability. Their impact will be analyzed in \cref{prop:SINR}.

\color{black}
\section{{Pilot Placement}}\label{sec:pilot}

\begin{figure}
    \centering
    \resizebox{0.85\linewidth}{!}{%
    \begin{tikzpicture}[everynode/.style={minimum size=.5cm-\pgflinewidth, inner sep=0pt},shading=rainbow]

\foreach \x in {-2, 6, 14}
{
\draw[dotted] (\x, 2) -- (\x, -6);
}
\node[above] at (2, 1.5) {\large TF TX$_1$};
\node[above] at (10, 1.5) {\large TF TX$_2$};

\fill[blue!25] (-1,1) rectangle (5,-5);
\draw[step=0.5cm,color=black] (-1,1) grid (5,-5);

\foreach \x in {-0.5, 0, 0.5, 1, 1.5, 2, 2.5, 3}{
\node[everynode,fill=white] at (\x+0.25,0.25) {\large $*$};
}
\foreach \y in {0.5, 0, -0.5, -1, -1.5, -2, -2.5, -3}{
\node[everynode,fill=white] at (-0.25,\y-0.25) {\large $*$};
}

\draw[->] (5,1) -- (5.25,1) node[right] {\large $n$};
\draw[->] (-1,-5) -- (-1,-5.25) node[below] {\large $m$};

\draw (-0.25, 1.5) -- (-0.25,1.25) node[right] {\large $n_{1(0)}$};
\draw[->] (-0.25, 1.25) -- (-0.25,1);
\draw (-1.25,0.25) -- (-1.5,0.25) node[above] {\large $m_{1(0)}$};
\draw[->] (-1.25,0.25) -- (-1,0.25);

\draw[->] (1.5,-5.25) -- (-0.5,-5.25);
\node[everynode,fill=white] at (1.75,-5.25) {\large $P_{\nu}$};
\draw[->] (2,-5.25) -- (3.5,-5.25);
\draw[->] (5.25,-2) -- (5.25,0.5);
\node[everynode,fill=white] at (5.25,-2.25) {\large $P_{\tau}$};
\draw[->] (5.25,-2.5) -- (5.25,-3.5);

\node[everynode, fill=white] at (2.75,-2.75) {\large $0$};
\node[everynode, fill=white] at (2.75-1,-2.75+1) {\large $0$};
\node[everynode, fill=white] at (2.75+1,-2.75+1) {\large $0$};
\node[everynode, fill=white] at (2.75-1,-2.75-1) {\large $0$};
\node[everynode, fill=white] at (2.75+1,-2.75-1) {\large $0$};

\fill[red!25] (-1+8,1) rectangle (5+8,-5);
\draw[step=0.5cm,color=black] (-1+8,1) grid (5+8,-5);

\foreach \x in {-0.5, 0, 0.5, 1, 1.5, 2, 2.5, 3}{
\node[everynode,fill=white] at (\x+0.25+8,0.25) {\large $\mathbf{0}$};
}
\foreach \y in {0.5, 0, -0.5, -1, -1.5, -2, -2.5, -3}{
\node[everynode,fill=white] at (-0.25+8,\y-0.25) {\large $\mathbf{0}$};
}

\draw[->] (5+8,1) -- (5.25+8,1) node[right] {\large $n$};
\draw[->] (-1+8,1) -- (-1+8,-5.25) node[below] {\large $m$};

\draw[->] (1.5+8,-5.25) -- (-0.5+8,-5.25);
\node[everynode,fill=white] at (1.75+8,-5.25) {\large $P_{\nu}$};
\draw[->] (2+8,-5.25) -- (3.5+8,-5.25);
\draw[->] (5.25+8,-2) -- (5.25+8,0.5);
\node[everynode,fill=white] at (5+8.25,-2.25) {\large $P_{\tau}$};
\draw[->] (5.25+8,-2.5) -- (5.25+8,-3.5);

\node[everynode, fill=white] at (2.75+8,-2.75) {\large $*$};
\node[everynode, fill=white] at (2.75-1+8,-2.75+1) {\large $*$};
\node[everynode, fill=white] at (2.75+1+8,-2.75+1) {\large $*$};
\node[everynode, fill=white] at (2.75-1+8,-2.75-1) {\large $*$};
\node[everynode, fill=white] at (2.75+1+8,-2.75-1) {\large $*$};

\end{tikzpicture}%
    }%
    \caption{TF-domain pilots (indicated by `$*$') and TF guard bins (indicated by `$0$') on TX$_1$ and TX$_2$. Each antenna has $20$ empty DD-domain bins.
    }%
    \label{OTFS_Pilot_Placement}%
\end{figure}

The exact TF I/O expression (\cref{thm:Y}) motivates the use of private TF bins for low-complexity scatterer estimation. In~\cite{wang2025ISACMIMO}, the concept rested on an approximate model with a subsymbol-dependent coefficient; \cref{thm:Y} is what enables the rigorous use of private bins for channel estimation. We briefly review this concept from~\cite{wang2025ISACMIMO}. A private TF bin $[n_p, m_p]$ for transmit antenna $p$ is created by assigning a pilot to antenna $p$ while enforcing zeros at the corresponding TF bin of all other transmit antennas. These zero-valued bins serve as TF guard bins, preserving the exclusive use of $[n_p, m_p]$ by antenna $p$ and maintaining pilot orthogonality across transmit antennas.

Introducing these pilots and guard bins, however, has a consequence for the invertibility of the TF-to-DD mapping. Since each TF bin is a linear combination of all DD-domain symbols through the ISFFT, reserving TF pilot and guard bins reduces the number of equations available for recovering the DD-domain symbols. To preserve the invertibility of the TF-to-DD mapping, each TF pilot/guard bin must therefore be paired with an empty DD bin, referred to as a DD guard bin. In~\cite{wang2025ISACMIMO}, the locations of these DD guard bins were not specified.

\color{black}
Here, we propose a principled method for selecting their locations. We enforce a rank condition based on Jacobi's complementary minor identity for unitary matrices~\cite{caragea2025principalminorsfouriermatrices}: the reduced ISFFT matrix is full rank if and only if the complementary submatrix indexed by the TF pilot/guard bins and the DD guard bins is full rank. Accordingly, the DD guard-bin locations are selected by checking this rank condition and are resampled whenever it is not satisfied.
\color{black}

The pilots are placed on private TF bins of a small number of antennas, forming line segments of length $P_\nu$ parallel to the time axis (time arm) and segments of length $P_\tau$ parallel to the frequency axis (frequency arm). 
In addition to the arm pilots, other pilots can be assigned to other antennas to exploit spatial diversity. We illustrate this pilot pattern in \cref{OTFS_Pilot_Placement} for the $N_t = 2$ case. 
A time arm and a frequency arm, with $P_\nu = P_\tau = 8$, are placed on TX$_1$, while $5$ pilots are placed on TX$_2$. On TX$_2$, the TF bins whose positions correspond to the arm pilots of TX$_1$ are set to $0$, serving as TF guard bins that maintain the orthogonality of the arm pilots. Similarly, TX$_1$ sets the TF bins whose positions correspond to the pilots of TX$_2$ to $0$.

Let $N_p$ denote the total number of pilots, which is primarily determined by the arm lengths and the total number of pilots on other antennas 
($P_\mathrm{random}$), \ie, $N_p \approx P_\tau + P_\nu + P_\mathrm{random}$. Because each TF pilot requires corresponding zeros on all other antennas, $N_p$ TF bins per antenna are reserved for pilots or guard bins. Furthermore, each TF pilot/guard bin requires one DD guard bin, for a total of $N_p$ DD guard bins per antenna. The values and positions of TF pilots, as well as the positions of DD guard bins, are known to the receiver via signaling.

\color{black}

Having established the pilot placement, we now characterize its performance.
The pilot is random (\eg, a QAM symbol) and independent of the non-pilot data symbols, so the interference in the pilot bins at the receiver is not only uncorrelated but also independent of the desired response, conditioned on the channel. Let $P_\mathrm{pilot}$ denote the pilot power, and assuming TF samples transmitted from different antennas are also uncorrelated, the SINR of the received pilot is given in \cref{prop:SINR}.

\begin{proposition}\label{prop:SINR}
The SINR of the received TF pilot bins is
\begin{align}\label{eq:SINR_simple}
    {\mathrm{SINR}}=\frac{P_{\rm Pilot}\sigma_{\beta}^2\sum_{j=0}^{J-1}\abs{\xi_j}^2}{N_t\sigma_{\beta}^2\sum_{j=0}^{J-1}\parens*{1-\abs{\xi_j}^2}+\sigma_{w}^2}.
\end{align}
\end{proposition}
The proof of \cref{prop:SINR} can be found in \cref{sec:appendix_I_short}. 

Accordingly, the desired SINR can be controlled by adjusting $P_{\rm Pilot}$, whose impact will be evaluated in \cref{sec:simulation}.


\color{black}

\section{Low-Complexity Coarse Channel Estimation} \label{sec:coarse_estimates}


The channel comprises $J$ scatterers. For the $j$-th scatterer, $\beta_j$ is the complex attenuation factor, and $(\theta_j,\varphi_j,\tau_j,\nu_j)$ are the AoA, AoD, delay, and Doppler shift, respectively.
Channel estimation, therefore, reduces to estimating these parameters.  
\textcolor{black}{In the following, the AoA of each scatterer is estimated based on all received bins, while the corresponding delay and Doppler are estimated based on the pilot bins. The AoD is then obtained from the estimated AoA and delay, using bistatic geometry.}

\underline{\textit{Coarse estimation of $\theta_j$ based on all received bins:}}
Ignoring $I_{n_c}[n,m]$ and $W_{n_c}[n,m]$, we proceed by rewriting \cref{eq:TFIO_rect_desired} as
\begin{align}\label{eq:TF_private}
 Y_{n_c}[n,m]
    &=\sum_{j=0}^{J-1} e^{-i2\pi n_c g_c\frac{\sin\theta_j}{\lambda}} A_j[n,m], 
\end{align}
where
\begin{flalign}
&A_j[n,m]
\!=\! \!\sum_{n_t=0}^{N_t-1}\! X_{n_t}[n,m] e^{-i2\pi n_t g_t\frac{\sin\varphi_j}{\lambda}} \beta_j H^j[n,m]\xi_j \!&&\! \label{eq:A_profile_1} 
\end{flalign}
Viewed as a function of $n_c$ for $n_c=0,1,\ldots,N_c-1$, \(Y_{n_c}[n,m]\) is the sum of $J$ discrete-time complex sinusoids, each with frequency $g_c{\sin(\theta_j)}/{\lambda}$ and complex amplitude $A_j[n,m]$.
The peak frequencies and corresponding amplitudes can be found using an $N_c$-point Discrete Fourier Transform (DFT). 
Based on the peak frequencies, the AoAs $\hat{\theta}_j$, serving as coarse estimates of $\theta_j$, can then be determined.
The DFT-based estimation can be applied to all TF-domain bins $[n,m]$, and the resulting angle estimates can then be averaged.


The resolution of AoA is limited by the aperture of receive array. The DFT peaks may not align with the ground truth.

\underline{\textit{Coarse estimation of} $\tau_j$, $\nu_j$ based on pilot bins:}
After the estimates $\hat{\theta}_{\textcolor{black}{{j}}}$, \textcolor{black}{for ${j}=0,1,\ldots,{j}-1$,} have been obtained, the corresponding values of $A_{\textcolor{black}{{j}}}[n, m]$s can be estimated from \cref{eq:TF_private} in the least-squares (LS) sense. 
In the following, we propose a novel, low-complexity approach to obtain coarse estimates of $\tau_j$, $\nu_j$ based on $A_{\textcolor{black}{{j}}}[n, m]$.
\color{black}
In the general case in which the angular resolution is insufficient to resolve the $N_{\textcolor{black}{{j}}}$ scatterers associated with the same $\hat{\theta}_{\textcolor{black}{{j}}}$, the ratio of $A_{\textcolor{black}{{j}}}[n,m]$, evaluated on pilot bin $[n_p,m_p]$ of antenna $p$, to the known pilot $X_p[n_p,m_p]$ can be expressed as follows:
\begin{multline}
    \frac{A_{\textcolor{black}{{j}}}[n_p,m_p]}{X_{p}[n_p,m_p]} \!=\!  \sum_{j_r=0}^{N_{\textcolor{black}{{j}}}-1} e^{-i2\pi p g_t \frac{\sin\varphi_{j_r}}{\lambda}} 
    \beta_{j_r} H^{j_r}[n_p,m_p]\xi_{j_r}, \label{A_profile_4}
\end{multline}
where $H^{j_r}[n_p,m_p]$ is given in \cref{eq:TFChannel}.
Here, we used the fact that pilot bins are privately assigned; thus, the summation over $N_t$ in \cref{eq:A_profile_1} reduces to the single term of antenna $p$.

\textcolor{black}{
The $P_{\nu}$ time arm (fixed frequency index, varying time index) collects samples of  \(\frac{{A_j}[n_p,m_p]}{X_{p}[n_p,m_p]}\)
whose phase evolution over time depends solely on linear Doppler shifts. Applying a 1D DFT along this arm yields coarse Doppler estimates.}
%
In particular,
evaluating \cref{A_profile_4} on the pilot time arm of antenna $p$, \ie, for a fixed $m_p$ and for $n_p=n_{p_0}, \ldots, n_{p_0}+P_{\nu}-1$, where $n_{p_0}$ is the starting index of the time arm, we have
\begin{align}
  \frac{A_{\textcolor{black}{{j}}}[n_p,m_p]}{X_p[n_p,m_p]}
  &= \sum_{j_r=0}^{N_{\textcolor{black}{{j}}}-1} G^t_{j_r}(p,m_p) e^{i 2\pi \bracks*{\frac{(k_{j_r}+\kappa_{j_r})n_p}{N}}}, \label{doppler-estimation}
\end{align}
where $G^t_{j_r}(p,m_p)$ is
\begin{align}
  e^{-i2\pi p g_t \frac{\sin\varphi_{j_r}}{\lambda}} \beta_{j_r} e^{-i2\pi\frac{(k_{j_r}+\kappa_{j_r})l_{j_r}}{NM}} e^{-i2\pi \frac{m_p l_{j_r}}{M}}\xi_{j_r}.
\end{align}
Viewed as a function of \(n_p\), \cref{doppler-estimation} is a finite segment of a sum of sinusoids with frequencies \(k_{j_r}+\kappa_{j_r}\), \(j_r=0,1,\ldots, N_{\textcolor{black}{{j}}}-1\). 

\textcolor{black}{
\Cref{doppler-estimation} relies on the subsymbol-independent coefficient established in \cref{thm:Y}, which makes \(A_{\textcolor{black}{{j}}}[n_p,m_p]\) a pure sinusoid in \(n_p\). In contrast, the TF-domain approximation in~\cite{wang2025ISACMIMO} yields sinusoids with \(n_p\)-dependent amplitudes from \cite[Eq.~(32)]{wang2025ISACMIMO}, precluding low-complexity estimation of \(k_{j_r}+\kappa_{j_r}\).
}

Similarly, evaluating \cref{A_profile_4} on the pilot frequency arm of antenna $p$, \ie, for a fixed $n_p$ and varying $m_p$, we have
\begin{align}
  \frac{A_{\textcolor{black}{{j}}}[n_p,m_p]}{X_p[n_p,m_p]}= \sum_{j_r=0}^{N_{\textcolor{black}{{j}}}-1} G^f_{j_r}(p,n_p) e^{\textcolor{black}{-}i 2\pi \bracks*{\frac{l_{j_r} m_p}{\textcolor{black}{M}}}}, \label{delay-estimation}
\end{align}
where $G^f_{j_r}(p,n_p)$ is
\begin{align}
  e^{-i2\pi p g_t \frac{\sin\varphi_{j_r}}{\lambda}} \beta_{j_r} e^{-i2\pi\frac{(k_{j_r}+\kappa_{j_r})l_{j_r}}{NM}} e^{i 2\pi \frac{(k_{j_r}+\kappa_{j_r})n_p}{N}} \xi_{j_r}.
\end{align}
Viewed as a function of $m_p$, \cref{delay-estimation} is a segment of the sum of sinusoids whose frequencies are $l_{j_r}$ for $j_r=0,1, \ldots, N_{\textcolor{black}{{j}}}-1$. 

The resolution of coarse estimates $\hat{k}_{j_r}+\hat{\kappa}_{j_r}$ (\ie $\hat{\nu}_{j_r}$) depends on $P_{\nu}$,
and the resolution of coarse estimates $\hat{l}_{j_r}$ (\ie $\hat{\tau}_{j_r}$) depends on $P_{\tau}$.
The maximum possible resolution is achieved when $P_{\nu}=N$ and $P_{\tau}=M$. At low resolution, the DFT peaks may not align with the ground truth.

\underline{\textit{Coarse estimation of} $\varphi_j$ using bistatic geometry:}
\textcolor{black}{
We assume two-dimensional (2D) propagation, where the transmitter, receiver, and scatterers lie in a common horizontal plane, and that the baseline distance between the transmitter and receiver is known via precalibration or deployment topology---a common bistatic assumption~\cite{willis2005bistatic}; however, direct Line-of-Sight (LoS) propagation is not required.}
After obtaining the coarse AoA and delay estimates $\hat{\theta}_j$ and $\hat{\tau}_j$, we can derive the coarse AoD estimates $\hat{\varphi}_j$.
From \cref{fig:bistaticdiagram}, the range estimates satisfy 
$c\hat{\tau}_j=\hat{R}_j=\hat{R}_{cj}+\hat{R}_{jt}$.
The LoS distance $R_{\rm LoS}$ \textcolor{black}{(not to be confused with the propagation path of a scatterer)} is either known a priori (for fixed base stations) or calculated from the signal's travel time, $\tau_{\rm LoS}$, \ie, $R_{\rm LoS} = c\tau_{\rm LoS}$.
Based on the law of cosines,
\begin{equation}\label{eq:LOC_1}
    (\hat{R}_j-\hat{R}_{cj})^2=\hat{R}_{jt}^2
    =R_{\rm LoS}^2+\hat{R}_{cj}^2-2R_{\rm LoS} \hat{R}_{cj}\cos(\hat{\theta}_j),
\end{equation}
and 
\begin{align}\label{eq:LOC_2}
    \hat{R}_{cj}^2 
    &= R_{\rm LoS}^2+\hat{R}_{jt}^2-2R_{\rm LoS}\hat{R}_{jt}\cos(\hat{\varphi}_j), 
 \end{align}
it holds that
\begin{align}\label{eq:bi-varphi}
    \hat{\varphi}_j
    &=\arccos\frac{R_{\rm LoS}^2+\hat{R}_{jt}^2-\hat{R}_{cj}^2}{2R_{\rm LoS}\hat{R}_{jt}}.
\end{align}

The resolution of $\hat{\varphi}_j$ depends on those of the coarse estimates $\hat{\theta}_j$ and $\hat{R}_j$; hence, $\hat{\varphi}_j$ is also coarse.

\begin{figure}
    \centering
    \resizebox{\columnwidth}{!}{%
    \begin{tikzpicture}
    \coordinate (origo) at (2,0.25);
    \coordinate (endgo) at (11,0.25);        
    \foreach \x in {0, 0.5, 1}{
    \node[above] at (\x,0) {\faSatelliteDish};
    }
    \draw (-0.5,0.1) -- (1.5,0.1) node[below,midway] {Transmitter};
    \foreach \x in {12, 12.5, 13}{
    \node[above, xscale=-1] at (\x,0) {\faSatelliteDish};
    }
    \draw (11.5,0.1) -- (13.5,0.1) node[below,midway,align=center] {Communication \ Receiver};
    \draw (2,0.25) -- (11,0.25) node[below,midway] (LoS) {LoS ($R_{\rm LoS}$)};
    \node[above,align=center] at (6.5,5) {$\nu_{2}$ \ \faTrain\ ($j=2$)};
    \draw[thick,->,orange!70!black] (2,0.25) -- (6.5,5) node[left,midway] (start) {$R_{2t}$};
    \draw[thick,->,ForestGreen] (6.5,5) -- (11,0.25) node[right,midway] (end) {$R_{c2}$};
    \pic [draw=red] {angle = LoS--origo--start};
    \draw (2.45,0.5) -- (2,1) node[left] {$(\varphi_{0},\varphi_{1},\varphi_{2})$};
    \pic [draw=red] {angle = end--endgo--LoS};
    \draw (10.55,0.5) -- (11,1) node[right] {$(\theta_{0},\theta_{1},\theta_{2})$};
    \node[above,align=center] at (6.5,3) {$\nu_{1}$ \ \faBus\ ($j=1$)};
    \draw[thick,->,orange!70!black] (2,0.25) -- (6.5,3) node[above,midway] {$R_{1t}$};
    \draw[thick,->,ForestGreen] (6.5,3) -- (11,0.25) node[above,midway] {$R_{c1}$};
    \node[above,align=center] at (6.5,1) {$\nu_{0}$ \ \faMotorcycle\ ($j=0$)};
    \draw[thick,->,orange!70!black] (2,0.25) -- (6.5,1) node[above,midway] {$R_{0t}$};
    \draw[thick,->,ForestGreen] (6.5,1) -- (11,0.25) node[above,midway] {$R_{c0}$};
\end{tikzpicture}
    }%
    \caption{Bistatic geometry with three scatterers.}
    \label{fig:bistaticdiagram}
\end{figure}

After acquiring coarse estimates, $(\hat{\theta}_j,\hat{\varphi}_j,\hat{\tau}_j,\hat{\nu}_j)$, the coarse path gain $\hat{\beta}_j$ can be obtained from the pilots on all receive antennas in the LS sense.
The resolution of those estimates is limited by the number of receive antennas and pilots.
In the next section, we refine these low-resolution estimates by constructing a virtual array and using SSR processing.


\textcolor{black}{
Of course, more sophisticated strategies can be employed to estimate the desired parameters in the presence of interference. For instance, SIC could exploit the explicit characterization of $I[n,m]$ (\cref{thm:Y}), estimating and subtracting each path's expected interference contribution before re-estimating the remaining paths. Alternatively, a minimum mean-square-error (MMSE) or maximum-likelihood (ML) estimator could incorporate the interference statistics directly into the estimation criterion, or subspace-based methods~\cite{kay1988modern} could be used. However, these approaches introduce additional computational complexity, while, based on our experience, offering only limited improvement in the final performance, which is obtained via the virtual array and SSR processing presented next.}



\section{Virtual Array and Sparse Signal Recovery}\label{sec:SSR}

\textcolor{black}{
The coarse estimates obtained in \cref{sec:coarse_estimates} are insufficient for accurate channel reconstruction due to their limited resolution. {Here, we} exploit the orthogonality of the proposed TF pilots to construct a virtual array and formulate an SSR problem to refine them, following the general approach of~\cite{wang2025ISACMIMO}. However, unlike the three-dimensional (3D) SSR problem of~\cite{wang2025ISACMIMO}, MIMO-OTFS CSI acquisition involves a four-dimensional (4D) parameter space $(\theta,\varphi,\tau,\nu)$, for which a direct dictionary discretization is computationally prohibitive and increases dictionary coherence. In this section, we develop a dimensionality-reduction framework that transforms the 4D problem into a sequence of much smaller SSRs by combining the coarse estimates, bistatic geometric constraints, and the separable structure of the steering vectors.
}

Based on the received signal on bin $[n_p,m_p]$ and the corresponding pilot $X_p[n_p,m_p]$, each receive antenna forms the ratio
$Y_{n_c}[n_p,m_p]/X_p[n_p,m_p]$.
These ratios are collected into the vector
$\mathbf r_p\in\mathbb C^{N_c\times1}$,
which is expressed as
\begin{align}
    \mathbf r_p
    = \mathbf\Phi_p(\theta_j,\varphi_j,\tau_j,\nu_j;n_p,m_p)\bm\beta,
    \label{17}
\end{align}
where
$\mathbf\Phi_p\in\mathbb C^{N_c\times J}$ is a matrix whose $(n_c,j)$-th element is
\begin{align}
    e^{-i2\pi n_c g_c\frac{\sin\theta_j}{\lambda}}
    e^{-i2\pi p g_t\frac{\sin\varphi_j}{\lambda}}
    H^j[n_p,m_p]\xi_j,
    \label{eq:CSI_element1}
\end{align}
and
$\bm\beta=[\beta_0,\beta_1,\ldots,\beta_{J-1}]^{\rm T}$.

By stacking the vectors corresponding to all pilot bins,
$\{\mathbf r_p\,|\,p=0,\ldots,N_p-1\}$,
into
$\mathbf r\in\mathbb C^{N_cN_p\times1}$,
we obtain
\begin{align}
    \mathbf r
    =\mathbf\Phi\bm\beta,
    \label{overcomplete}
\end{align}
where each column of
$\mathbf\Phi\in\mathbb C^{N_cN_p\times J}$
is parameterized by
$(\theta_j,\varphi_j,\tau_j,\nu_j)$.
\Cref{overcomplete} forms a virtual array that provides an
$N_p$-fold increase in angular resolution compared with the physical receive array.

The virtual array can be reformulated as an overcomplete sparse representation by discretizing the parameter space
$(\theta,\varphi,\tau,\nu)$ into
$(G_\theta,G_\varphi,G_\tau,G_\nu)$ grid points,
yielding
\begin{align}
    \mathbf r
    =\tilde{\mathbf\Phi}\tilde{\bm\beta},
    \label{eq:CSI_mtx3}
\end{align}
where
$\tilde{\mathbf\Phi}\in
\mathbb C^{N_cN_p\times
G_\theta G_\varphi G_\tau G_\nu}$
is the dictionary matrix and
$\tilde{\bm\beta}$ is a sparse vector whose nonzero entries identify the scatterers.
The channel parameters can then be estimated by solving the SSR problem using orthogonal matching pursuit (OMP)~\cite{gomez-cuba2021CompressedSensing}.

The performance of the SSR algorithm depends critically on the size of the dictionary matrix $\tilde{\mathbf{\Phi}}$, whose columns (atoms) correspond to candidate channel parameters. Although SSR can solve underdetermined systems, excessively large dictionaries increase computational complexity and dictionary coherence, leading to incorrect atom selection. A straightforward discretization of the four-dimensional parameter space $(\theta,\varphi,\tau,\nu)$ produces $G_\theta G_\varphi G_\tau G_\nu$ atoms. 
For example, $G_\theta\times G_\varphi\times G_\tau\times G_\nu=5\times15\times5\times10=3750$,
and finer discretizations increase this number rapidly. Consequently, directly solving the 4D SSR problem does not scale to large systems.

\color{black}

Here we propose a
dimensionality-reduction framework for the SSR problem. We adopt the coarse-to-fine search strategy developed in~\cite{wang2025ISACMIMO} to initialize the search around the coarse estimates obtained in \cref{sec:coarse_estimates}. Building on this initialization, we introduce two additional reductions specific to the proposed MIMO-OTFS channel estimation framework. First, we exploit the bistatic geometry to eliminate the AoD dimension, thereby reducing the original 4D search to a 3D one. Second, we exploit the separable structure of the steering vectors to decompose the remaining 3D SSR problem into two sequential lower-dimensional sparse recovery problems. Finally, binary refinement is applied within the reduced search space to achieve high-resolution estimates. These reductions substantially decrease the computational complexity while preserving estimation accuracy.

\begin{algorithm}
\caption{Two-Stage OMP With Binary Refinement}
\label{alg:omp}
\begin{algorithmic}[1]
    \Require Virtual array $\mathbf{r}$, initial step sizes $\bm{\delta}=(\delta_{\theta}, \delta_{\tau}, \delta_{\nu})$, final step sizes $\bm{\delta}_{\min}=(0.1^{\circ}, \Delta\tau, 0.1\Delta\nu)$, noise power $\textcolor{black}{\sigma_{w}^2}$, coarse estimates $\{\hat\theta_j, \hat\tau_j, \hat\nu_j\}_{j=0}^{J-1}$, grid points $(G_{\theta}, G_{\tau}, G_{\nu})$.
    \Ensure Refined SSR estimates $\{\hat\theta_j, \hat\varphi_j, \hat\tau_j, \hat\nu_j, \hat\beta_j\}_{j=0}^{\textcolor{black}{J-1}}$.    
    \While{any($\bm{\delta} > \bm{\delta}_{\min}$)} 
        \State Initialize centers: $(\bar\theta_j, \bar{\tau}_j, \bar\nu_j) \!\leftarrow\! \{\hat\theta_j, \hat\tau_j, \hat\nu_j\}_{j=0}^{J-1}$    
        \State Initialize residual: $\mathbf{r}_{\text{res}} \leftarrow \mathbf{r}$, Active Set $\mathcal{A} \leftarrow \emptyset$, $j^*=0$
        \State \textit{// Two-Stage OMP}
        \While{$j^*<\textcolor{black}{J}$} 
            \State \textbf{Stage 1} (refine delay-Doppler, fixed angles):
            \State \hspace{\algorithmicindent} Construct the dictionary $\tilde{\mathbf{\Phi}}_{\tau,\nu}$ with fixed $(\bar\theta_j)$, grid points $(G_{\tau},G_{\nu})$, step sizes $(\delta_{\tau},\delta_{\nu})$, and derived $\tilde{\varphi}_j$.
            \State \hspace{\algorithmicindent} $(\hat\tau_{j^*}, \hat\nu_{j^*}) \leftarrow \arg\max_{(\tilde\tau, \tilde\nu)} |\tilde{\mathbf{\Phi}}_{\tau,\nu}\hmt \mathbf{r}_{\text{res}}|$
            \State \textbf{Stage 2} (refine angles, fixed delay-Doppler):
            \State \hspace{\algorithmicindent} Construct the dictionary $\tilde{\mathbf{\Phi}}_{\theta,\varphi}$ with $(\hat\tau_{j^*}, \hat\nu_{j^*})$, grid points $(G_{\theta})$, step sizes $(\delta_{\theta})$, and derived $\tilde{\varphi}_j$.
            \State \hspace{\algorithmicindent} $(\hat\theta_{j^*}, \hat\varphi_{j^*}) \leftarrow \arg\max_{(\tilde\theta, \tilde\varphi)} |\tilde{\mathbf{\Phi}}_{\theta,\varphi}\hmt \mathbf{r}_{\text{res}}|$
                
            \State Update active set: $\mathcal{A} \leftarrow \mathcal{A} \cup \{(\hat\theta_{j^*}, \hat\varphi_{j^*}, \hat\tau_{j^*}, \hat\nu_{j^*})\}$
            \State Construct dictionary $\mPhi_{\mathcal{A}}$ using parameters of all $\mathcal{A}$
            \State Solve gains: $\hat{\bm{\beta}}_{\mathcal{A}} = (\mPhi_{\mathcal{A}}^H \mPhi_{\mathcal{A}} + \textcolor{black}{\sigma_{w}^2} \mI)^{-1} \mPhi_{\mathcal{A}}^H \mathbf{r}$
            \State Update residual: $\mathbf{r}_{\text{res}} = \mathbf{r} - \mPhi_{\mathcal{A}} \hat{\bm{\beta}}_{\mathcal{A}}$
            \State Update $j^* \leftarrow j^*+1$
        \EndWhile        
        \State \textit{// Halve step sizes for binary refinement}
        \State $\bm{\delta} \leftarrow \max( \bm{\delta} / 2, \bm{\delta}_{\min} )$
    \EndWhile
\end{algorithmic}
\end{algorithm}

\textit{Initialization (coarse localization):}
The coarse estimates are used to restrict the search space of
$\{\theta,\tau,\nu\}$ to local neighborhoods around each estimate,
\begin{align}
\tilde{x}_j\in
\left[
\hat{x}_j-{G_x\delta_x}/{2},
\hat{x}_j+{G_x\delta_x}/{2}
\right],
\qquad
x\in\{\theta,\tau,\nu\},
\end{align}
where $\delta_x$ denotes the discretization step size. The initial search window should satisfy
\begin{align}
{G_x\delta_x}/{2}>\epsilon_x,
\qquad
x\in\{\theta,\tau,\nu\},
\end{align}
where
\begin{align}
\epsilon_\theta=\frac{\pi}{2N_c},\qquad
\epsilon_\tau=\frac{\tau_{\max}}{2P_\tau},\qquad
\epsilon_\nu=\frac{\nu_{\max}}{4P_\nu},
\end{align}
corresponding to the spacing between adjacent DFT bins in angle, delay, and Doppler.

\textit{Reduction 1 (Geometry-based elimination of the AoD dimension):}
For every candidate pair $(\tilde{\theta}_j,\tilde{\tau}_j)$, the total path length $\tilde{R}_j=c\tilde{\tau}_j$ together with $\tilde{\theta}_j$ uniquely determines $\tilde{R}_{cj}$ and $\tilde{R}_{jt}$ through~\cref{eq:LOC_1}, with $\tilde{R}_{jt}=\tilde{R}_j-\tilde{R}_{cj}$. The corresponding AoD is then obtained from~\cref{eq:bi-varphi}. As a result, the dictionary is constructed over the reduced parameter space $\{\theta,\tau,\nu\}$ rather than the original four-dimensional space.

\textit{Reduction 2 (Alternating lower-dimensional sparse recovery):}
We further reduce the complexity by exploiting the separable structure of the dictionary atoms in~\cref{eq:CSI_element1}. Instead of searching jointly over the remaining three-dimensional parameter space, we alternate between two lower-dimensional sparse recovery problems. In the first stage, $\theta$ is fixed while $(\tau,\nu)$ are refined. In the second stage, $(\tau,\nu)$ are fixed while $\theta$ is refined. The AoD is updated geometrically in both stages. This decomposition reduces the dictionary size in each OMP iteration from
$G_\theta G_\tau G_\nu$
atoms to two much smaller dictionaries containing
$G_\tau G_\nu$
and
$G_\theta$
atoms, respectively.

\textit{Binary refinement:}
To obtain high-resolution estimates, we apply the binary refinement strategy of~\cite{gomez-cuba2021CompressedSensing}, progressively shrinking the search region after each iteration. The refinement of the Doppler parameter continues to fractional-grid spacing, enabling accurate estimation of fractional Doppler shifts. The complete algorithm is summarized in \cref{alg:omp}.

\color{black}

\section{Recovering the Communication Information}\label{sec:modified_sfft}

The received DD-domain symbol on bin $[k,l]$ at antenna $n_c$, obtained by generalizing \cref{eq:DDIO_rect} to a MIMO system, is
\begin{align}
    y_{n_c}[k,l]
    &= \sum_{j=0}^{J-1} e^{-i2\pi n_cg_c \frac{\sin\theta_j}{\lambda}} 
    \sum_{n_t=0}^{N_t-1} e^{-i2\pi n_tg_t \frac{\sin\varphi_j}{\lambda}} \\
    &\times \sum_{\substack{q=k_j-\\ (N-1)}}^{k_j} x_{n_t}[[k-(k_j-q)]_N,[l-l_j]_M] 
    \\
    &\times
    \beta_je^{-i2\pi\frac{(k_j+\kappa_j) l_j}{NM}} \frac{1}{N} \alpha_j[k,l,q] +w_{n_c}[k,l].
\end{align}

The vectorized MIMO-OTFS I/O can be expressed as~\cite{kollengoderamachandran2018MIMOOTFSHighDoppler}
\begin{align}\label{eq:TFIO}
    \mathbf{y}_{\text{MIMO}}
    &=  \mathbf{h}_{\text{MIMO}}\mathbf{x}_{\text{MIMO}} 
    +\mathbf{w}_{\text{MIMO}}
    ,
\end{align}
where $\mathbf{y}_{\text{MIMO}}\in\C^{N_cNM}$,  \(\mathbf{x}_{\text{MIMO}}\in\C^{N_tNM}\), and $\mathbf{w}_{\text{MIMO}}$ 
contain the received DD-domain symbols, transmitted DD-domain symbols, and AWGN, respectively, for all antennas, with the symbols of each antenna ordered row-wise; $\mathbf{h}_{\text{MIMO}}\in\C^{N_cNM\times N_tNM}$ is the channel matrix, which equals~\cite{srivastava2022DelayDopplerAngular}
\begin{multline}\label{eq:effective_channel}
    \mathbf{h}_{\text{MIMO}}
    = \sum_{j=0}^{J-1}\beta_j(\mathbf{a}_{c}(\theta_j)\mathbf{a}_{t}\hmt(\varphi_j)) \otimes \\
    (\mathbf{F}_{\text{N}}\otimes\mathbf{I}_{\text{M}})
    (\Pi^{l_j}\Delta^{k_j+\kappa_j})
    (\mathbf{F}_{\text{N}}\hmt\otimes\mathbf{I}_{\text{M}}),
\end{multline}
where $\otimes$ is the Kronecker product, $\mathbf{F}_{\text{N}}$ is the DFT matrix of order $N$, $\mathbf{I}_{\text{M}}$ is the identity matrix of order $M$, $\Pi$ is the forward cyclic shift permutation matrix of size $NM\times NM$, and $\Delta=\diag\bracks*{z^0,z^1,\ldots,z^{NM-1}}$ with $z=e^{i2\pi\frac{1}{NM}}$.


After refining $(\theta_j,\varphi_j,\tau_j,\nu_j)$ and $\beta_j$, the CSI $\hat{\mathbf h}_{\rm MIMO}$ is constructed via \cref{eq:effective_channel}. Symbol detection begins by subtracting the pilot contribution from the received signal, \ie,
\begin{align}\label{eq:received_data}
\mathbf{y}_{\text{MIMO, data}} = \mathbf{y}_{\text{MIMO}} - \hat{\mathbf{h}}_{\text{MIMO}}\mathbf{x}_{\text{MIMO, pilot}},
\end{align}
where $\mathbf{x}_{\text{MIMO, pilot}}$ is the SFFT of TF pilots $\mathbf{X}_{\text{MIMO, pilot}}$.
Despite the loss of information from TF guard zeros and pilots, the DD guard bins guarantee a sufficient number of linear equations for LMMSE equalization to recover the transmitted symbols:
\begin{align*}
\hat{\mathbf{x}}_{\text{MIMO, data}} = (\hat{\mathbf{h}}_{\text{MIMO}}\hmt \hat{\mathbf{h}}_{\text{MIMO}} + \textcolor{black}{\sigma_{w}^2} \mathbf{I})\inv \hat{\mathbf{h}}_{\text{MIMO}}\hmt \mathbf{y}_{\text{MIMO, data}}. 
\end{align*}
Because $\hat{\mathbf{h}}_{\text{MIMO}}$ is highly sparse, this equation can be solved efficiently without direct matrix inversion using the least-squares with QR factorization (LSQR). 
Finally, $\hat{\mathbf{x}}_{\text{MIMO, data}}$ sets DD guard bins positions to zero, completing symbol detection.

\section{Pilot Overhead Comparison}\label{sec:overhead}

The overhead of prior methods that use  non-overlapped DD-pilot designs~\cite{raviteja2019EmbeddedPilotAided}-\cite{kollengoderamachandran2018MIMOOTFSHighDoppler} is
$\eta=N_t[(N_t+1)l_{\tau_{\text{max}}}+N_t][4k_{\nu_{\text{max}}}+1]/(N_tNM)$, which depends on the number of transmit antennas $N_t$. This dependence limits the scalability of those methods to large arrays.
This overhead is also subject to the maximum Doppler shift and delay.
Because such prior information is not readily available, a larger guard region may be required, further increasing the overhead.

For prior methods that use overlapped DD-pilots with shared guard regions~\cite{shen2019ChannelEstimation}-\cite{liang2024TwoDimensionalDelayDoppler},
the overhead is reduced to $\eta=N_t[2l_{\tau_{\text{max}}}+1][4k_{\nu_{\text{max}}}+1]/(N_tNM)$, which does not depend on $N_t$ after simplification.
However, the overhead still depends on the maximum Doppler and delay.
Furthermore, this overlapped design causes pilot pollution, degrading performance as the number of transmit antennas increases.

Our proposed method reduces the number of information-bearing symbols by \( N_t \) for each TF pilot, resulting in a total pilot overhead of \(\eta = {N_t N_p}/{(N_t N M)}\), where $N_p\approx P_{\tau}+P_{\nu}+P_{\rm random}$. 
This overhead is independent of the number of transmit antennas, ensuring scalability for large MIMO arrays.
Unlike existing methods (\cite{raviteja2019EmbeddedPilotAided}-\cite{kollengoderamachandran2018MIMOOTFSHighDoppler}), our overhead does not depend on channel parameters, thereby avoiding the need for prior channel information and further reducing overhead.
Our design achieves orthogonality at the receiver, facilitating accurate estimation with minimal pilots and less pilot pollution.

\Cref{tab:overhead} compares the pilot overhead of different methods. As shown, the proposed method has the lowest pilot overhead and is independent of $N_t$ and channel parameters. 
\textcolor{black}{Here, we consider $P_{\rm random}=0$ for simplicity. As shown in \cref{sec:simulation}, the proposed method can achieve good performance without additional pilots and can achieve substantial improvement with a small $P_{\rm random}$.}
The overhead of methods in~\cite{shen2019ChannelEstimation}-\cite{liang2024TwoDimensionalDelayDoppler} is independent of $N_t$ but depends on the channel parameters, while the overhead of the methods in~\cite{raviteja2019EmbeddedPilotAided}-\cite{kollengoderamachandran2018MIMOOTFSHighDoppler} depends on both $N_t$ and the channel parameters.

\begin{table}
\centering
\caption{Overhead comparison ($M=512$, $N=128$)}
\label{tab:overhead}
\resizebox{\columnwidth}{!}{%
\begin{tabular}{|l|ll|l|l|}
\hline
Pilot Placement &
  \multicolumn{2}{c|}{Overhead ($\eta$)} &
  \begin{tabular}[c]{@{}l@{}}$l_{\tau_{\max}}\!=\!10$\\ $k_{\nu_{\max}}\!=\!6$\end{tabular} &
  \begin{tabular}[c]{@{}l@{}}$l_{\tau_{\max}}\!=\!15$\\ $k_{\nu_{\max}}\!=\!10$\end{tabular} \\ \hline
\multirow{2}{*}{\begin{tabular}[c]{@{}l@{}} Non-overlapped \\ DD pilot~\cite{raviteja2019EmbeddedPilotAided}-\cite{kollengoderamachandran2018MIMOOTFSHighDoppler} \end{tabular}} &
  \multicolumn{1}{l|}{\multirow{2}{*}{$\frac{[(N_t\!+\!1)l_{\tau_{\text{max}}}\!+\!N_t][4k_{\nu_{\text{max}}}\!+\!1]}{NM}$}} &
  $N_t\!=\!4$ &
  $2.1\%$ &
  $4.9\%$ \\ \cline{3-5} 
 & \multicolumn{1}{l|}{} & $N_t\!=\!8$               & $3.7\%$  & $8.9\%$  \\ \hline
\multirow{2}{*}{\begin{tabular}[c]{@{}l@{}} Overlapped \\ DD Pilot~\cite{shen2019ChannelEstimation}-\cite{liang2024TwoDimensionalDelayDoppler} \end{tabular}} &
  \multicolumn{2}{l|}{\multirow{2}{*}{$ \frac{[2l_{\tau_{\text{max}}}\!+\!1][4k_{\nu_{\text{max}}}\!+\!1]}{NM}$ (Independent of $N_t$)}} &
  \multirow{2}{*}{$0.8\%$} &
  \multirow{2}{*}{$1.9\%$} \\
 & \multicolumn{2}{l|}{}                         &          &          \\ \hline
\multirow{2}{*}{Proposed TF Pilot} &
  \multicolumn{1}{l|}{\multirow{2}{*}{$\frac{N_p}{NM}$ (Independent of $N_t$)}} &
  $P_{\tau}\!=\!P_{\nu}\!=\!64$ &
  \textcolor{black}{$0.2\%$} &
  \textcolor{black}{$0.2\%$} \\ \cline{3-5} 
 & \multicolumn{1}{l|}{} & $P_{\tau}\!=\!P_{\nu}\!=\!128$ & \textcolor{black}{$0.4\%$} & \textcolor{black}{$0.4\%$} \\ \hline
\end{tabular}%
}
\end{table}
\section{Simulation Results}\label{sec:simulation}


In this section, we present simulation results to demonstrate the performance of the proposed system. 
\textcolor{black}{
We consider the system operating at a $4$~GHz carrier frequency and a $30$~kHz subcarrier spacing, consistent with the 5G NR FR1 protocol~\cite{vook20185gnr}, and with a grid size $M\times N=512\times 128$, consistent with the existing literature~\cite{raviteja2018InterferenceCancellation,shen2019ChannelEstimation, liang2024TwoDimensionalDelayDoppler}.}
The antenna spacing is half the wavelength, the constellation is QPSK, $N_t\times N_c=4\times16$, and the pilot power is scaled to the average transmit power.


The scatterers lie in a 2D plane. 
The transmitter is located at the origin $(0, 0)$, and the communication receiver is positioned at $(100, 0)$, establishing a LoS distance of $R_{\text{LoS}} = 100$ m. 
The scatterers are generated randomly;
their delay taps $l_{cj} = R_{cj}/(c\Delta \tau)$ and $l_{jt} = R_{jt}/(c\Delta \tau)$ are integers drawn from $\{1, 2, \dots, 8\}$, and their AoAs $\theta_j$ and AoDs $\varphi_j$ are restricted to $[-60^\circ, 60^\circ]$.
Note that \(\varphi_j\) is set based on the other parameters as shown in \cref{eq:LOC_1,eq:LOC_2,eq:bi-varphi}. 
To model mobility, the scatterers are assigned speeds ranging from $65$ m/s to $130$ m/s, corresponding to Doppler indices $k_j+\kappa_j=v_j/(\lambda\Delta\nu)$ in the range of \textcolor{black}{$\pm[3.7,7.4]$} relative to the communication receiver, directed along $\theta_j$. 
These speeds are representative of high-mobility targets such as high-speed trains or fast-moving UAVs~\cite{itu2020minrequirements}. 
\textcolor{black}{The effective power retained in each path $\abs*{\xi_j}^2\in(0.929,0.989)$.}
Under such conditions, conventional OFDM systems experience significant performance degradation due to the severe Doppler effects. 
The path gains $\beta_j$ are drawn from a unit-variance complex Gaussian distribution, \ie $\sigma_{\beta}^2=1$.
Unless otherwise stated, the number of scatterers is $J=4$. 

The channel estimation performance is evaluated in terms of the normalized mean-squared error (NMSE) in [dB], \ie, 
\({\rm NMSE}=10\log_{10}{\norm{\hat{\mathbf{h}}_{\rm MIMO}-\mathbf{h}_{\rm MIMO}}_{\rm F}^2}/{\norm{\mathbf{h}_{\rm MIMO}}_{\rm F}^2}\)
where $\hat{\mathbf{h}}_{\rm MIMO}$
is the estimated CSI, obtained based on    
\cref{eq:effective_channel} using the estimated values of $\beta_j$ and $(\theta_j,\varphi_j,\tau_j,\nu_j)$;   $\mathbf{h}_{\rm MIMO}$ is the true CSI, again obtained based on \cref{eq:effective_channel} using the true value of those parameters; and $\norm{\cdot}_{\rm F}$ is the Frobenius norm.
For this analysis, we define the signal-to-noise ratio (SNR) as
\({\rm SNR} \triangleq {P_{\text{Signal}}}/{P_{\text{Noise}}} = {\E{||\mathbf{h}_{\text{MIMO}}\mathbf{x}_{\text{MIMO}}||^2}}/{\textcolor{black}{\sigma_{w}^2}},\)
where $P_{\text{Signal}}$ is the average received signal power and $P_{\text{Noise}} = \textcolor{black}{\sigma_{w}^2}$ is the average power of the additive white noise.
\textcolor{black}{The corresponding SINR can be derived from \cref{prop:SINR}.
For example, it increases from approximately \(4.6 \sim 5.8\) dB at $0$~dB SNR to \(10.7\sim17.2\) dB at $15$~dB and \(11.2\sim19.4\) dB at $30$~dB, revealing a SINR floor at high-SNR imposed by the residual path-wise interference.}



During simulation, each transmit antenna multiplexes data symbols on the DD grid, leaving $N_p$ randomly selected DD guard bins empty. 
After converting the DD-domain symbols into the TF domain, the pilots and TF guard bins are injected as described in \cref{sec:pilot}. 
The frequency arm of length $P_{\tau}$ is located at $[n,m]=[64, 1:P_{\tau}]$ on the TX$_1$, while the time arm of length $P_{\nu}$ is located at $[n,m]=[1:P_{\nu},256]$ on TX$_2$.
Additionally, $P_{\rm random}$ pilots are randomly injected into other antennas.
\textcolor{black}{By default, $P_{\tau}=P_{\nu}=64$ and $P_{\rm random}=16$, with $8$ pilots  allocated to each of $\mathrm{TX}_3$ and $\mathrm{TX}_4$.}
The transmitted time-domain signal is constructed as explained in \cref{eq:ISFFT,eq:Heisenberg} of \cref{sec:MIMO_OTFS}.
After passing through the channel, the signal is received and brought to the TF domain (\cref{thm:Y}).
Coarse channel parameter estimates are obtained as described in \cref{sec:coarse_estimates}.
\textcolor{black}{
Since accurate estimates of the noise power are not assumed to be available, peak frequency detection using techniques such as the generalized likelihood ratio test (GLRT) or constant false-alarm rate (CFAR) detection struggles to locate weak scatterers. Instead, we employ prominence-based peak detection \cite{BibEntry2026Feb}, which operates without prior noise statistics, and combine it with successive interference cancellation (SIC) to mitigate masking effects and reveal weak targets.
A scatterer can be revealed provided that its coarse estimates are separable in at least one of the AoA, Doppler, or delay domains. Increasing the corresponding resolutions \(N_c\), \(P_{\nu}\), and \(P_{\tau}\), therefore, improves path resolvability and reduces the NMSE. Since the coarse stage resolves all path components above the effective noise floor, the residual path energy is absorbed into the noise term, and the scatterers identified at this stage are subsequently refined by SSR
(\cref{sec:SSR}) based on the orthogonal received TF pilots.}
An SSR problem is formulated using a compactly constructed dictionary matrix based on the coarse estimates.
The number of grid points for each scatterer is $G_{\theta}=11$, 
$G_{\tau}=11$, and $G_{\nu}=21$.
The elements in $\tilde{\mathbf \Phi}$ are taken as described in \cref{sec:SSR}.
Upon solving the SSR problem, we obtain the refined estimates.

\subsubsection{Coarse Estimation Performance}

We first evaluate the performance of coarse channel parameter estimation for $(\theta, \tau, \nu)$ using a $500$-trial Monte Carlo simulation. 
In each trial, the scatterers' parameters are fixed according to \cref{tab:channel_parameters}, while the complex gains are randomly generated.
The results for $5$~dB SNR under two different pilot configurations are presented in \cref{tab:coarse_estimates}. 
As illustrated, the coarse estimates for $\theta_j$ remain invariant to the number of pilots, as their accuracy depends primarily on $\textcolor{black}{N_c}$. 
Conversely, for most scatterers (\eg, 1, 2, and 4), the standard deviation increases when fewer pilots are used. 
This variation stems from the inherent resolution limits of the DFT, as discussed in \cref{sec:coarse_estimates}.
Despite this, the mean values of the coarse estimates remain very close to the ground truth. 
Consequently, these estimates establish a compact and reliable search space for the SSR problem. 

\begin{table}
    \centering
    \caption{Example scatterer parameters}
    \label{tab:channel_parameters}
    \resizebox{\columnwidth}{!}{%
    \begin{tabular}{|c|c|c|c|c|c|c|}
    \hline
    $j$ & Coordinates      & $\theta_j^\circ$ & $l_j = l_{cj}+l_{jt}$ & $k_j+\kappa_j$ & $\varphi_j^\circ$ & $\abs{\xi_j}^2$ \\ \hline \hline
    $1$ & $(50.0, -30.5)$   & $-31.4$          & $ 8=4+4$          & $-4-0.2$  & $-31.4$  & 0.97  \\ \hline
    $2$ & $(59.6,  42.4)$   & $46.4$           & $9=4+5$           & $5+0.4$   & $35.4$  & 0.96  \\ \hline
    $3$ & $(50.0,  -53.5)$  & $-46.9$          & $10=5+5$          & $4+0.1$   & $-46.9$ & 0.96  \\ \hline 
    $4$ & $(72.5, 10.0)$    & $20.1$           & $7=2+5$           & $-3-0.3$  & $7.9$    & 0.97  \\ \hline
    \end{tabular}
    }
\end{table}

\begin{table}[]
\centering
\caption{Coarse-estimation performance at ${\rm SNR}=5{\rm ~dB}$}
\label{tab:coarse_estimates}
\resizebox{\columnwidth}{!}{%
\begin{tabular}{|c|cccccccc|}
\hline
\multirow{2}{*}{$j$} &
  \multicolumn{2}{c|}{$\hat{\theta}_j$} &
  \multicolumn{2}{c|}{$\hat{l}_j$} &
  \multicolumn{2}{c|}{$\hat{k}_j+\hat{\kappa}_j$} &
  \multicolumn{2}{c|}{$\hat{\varphi}_j$} \\ \cline{2-9} 
 &
  \multicolumn{1}{c|}{mean} &
  \multicolumn{1}{c|}{std.} &
  \multicolumn{1}{c|}{mean} &
  \multicolumn{1}{c|}{std.} &
  \multicolumn{1}{c|}{mean} &
  \multicolumn{1}{c|}{std.} &
  \multicolumn{1}{c|}{mean} &
  std. \\ \hline \hline
 &
  \multicolumn{8}{c|}{$P_{\tau}=P_{\nu}=32$} \\ \hline
$1$ &
  \multicolumn{1}{c|}{$-32.02$} &
  \multicolumn{1}{c|}{$0.39$} &
  \multicolumn{1}{c|}{$8.01$} &
  \multicolumn{1}{c|}{$0.14$} &
  \multicolumn{1}{c|}{$-4.21$} &
  \multicolumn{1}{c|}{$0.05$} &
  \multicolumn{1}{c|}{$-30.92$} &
  $8.35$ \\ \hline
$2$ &
  \multicolumn{1}{c|}{$46.00$} &
  \multicolumn{1}{c|}{$0.00$} &
  \multicolumn{1}{c|}{$9.03$} &
  \multicolumn{1}{c|}{$0.91$} &
  \multicolumn{1}{c|}{$5.40$} &
  \multicolumn{1}{c|}{$0.02$} &
  \multicolumn{1}{c|}{$35.83$} &
  $4.89$ \\ \hline
$3$ &
  \multicolumn{1}{c|}{$-46.05$} &
  \multicolumn{1}{c|}{$1.61$} &
  \multicolumn{1}{c|}{$10.00$} &
  \multicolumn{1}{c|}{$0.18$} &
  \multicolumn{1}{c|}{$4.06$} &
  \multicolumn{1}{c|}{$0.62$} &
  \multicolumn{1}{c|}{$-48.68$} &
  $3.14$ \\ \hline
$4$ &
  \multicolumn{1}{c|}{$19.63$} &
  \multicolumn{1}{c|}{$4.62$} &
  \multicolumn{1}{c|}{$7.04$} &
  \multicolumn{1}{c|}{$0.22$} &
  \multicolumn{1}{c|}{$-3.31$} &
  \multicolumn{1}{c|}{$0.10$} &
  \multicolumn{1}{c|}{$11.41$} &
  $11.69$ \\ \hline \hline
 &
  \multicolumn{8}{c|}{$P_{\tau}=P_{\nu}=64$} \\ \hline
$1$ &
  \multicolumn{1}{c|}{$-32.03$} &
  \multicolumn{1}{c|}{$0.37$} &
  \multicolumn{1}{c|}{$8.00$} &
  \multicolumn{1}{c|}{$0.00$} &
  \multicolumn{1}{c|}{$-4.20$} &
  \multicolumn{1}{c|}{$0.03$} &
  \multicolumn{1}{c|}{$-30.81$} &
  $3.06$ \\ \hline
$2$ &
  \multicolumn{1}{c|}{$46.00$} &
  \multicolumn{1}{c|}{$0.00$} &
  \multicolumn{1}{c|}{$9.00$} &
  \multicolumn{1}{c|}{$0.00$} &
  \multicolumn{1}{c|}{$5.40$} &
  \multicolumn{1}{c|}{$0.00$} &
  \multicolumn{1}{c|}{$35.39$} &
  $5.47$ \\ \hline
$3$ &
  \multicolumn{1}{c|}{$-46.06$} &
  \multicolumn{1}{c|}{$1.62$} &
  \multicolumn{1}{c|}{$9.99$} &
  \multicolumn{1}{c|}{$0.15$} &
  \multicolumn{1}{c|}{$4.05$} &
  \multicolumn{1}{c|}{$0.62$} &
  \multicolumn{1}{c|}{$-47.70$} &
  $0.91$ \\ \hline
$4$ &
  \multicolumn{1}{c|}{$19.63$} &
  \multicolumn{1}{c|}{$4.62$} &
  \multicolumn{1}{c|}{$7.01$} &
  \multicolumn{1}{c|}{$0.10$} &
  \multicolumn{1}{c|}{$-3.31$} &
  \multicolumn{1}{c|}{$0.09$} &
  \multicolumn{1}{c|}{$7.9$} &
  $0.00$ \\ \hline
\end{tabular}%
}
\end{table}

\subsubsection{NMSE Performance}


\begin{figure*}
    \centering
    \subfloat[]{%
        \includegraphics[width=0.5\textwidth]
        {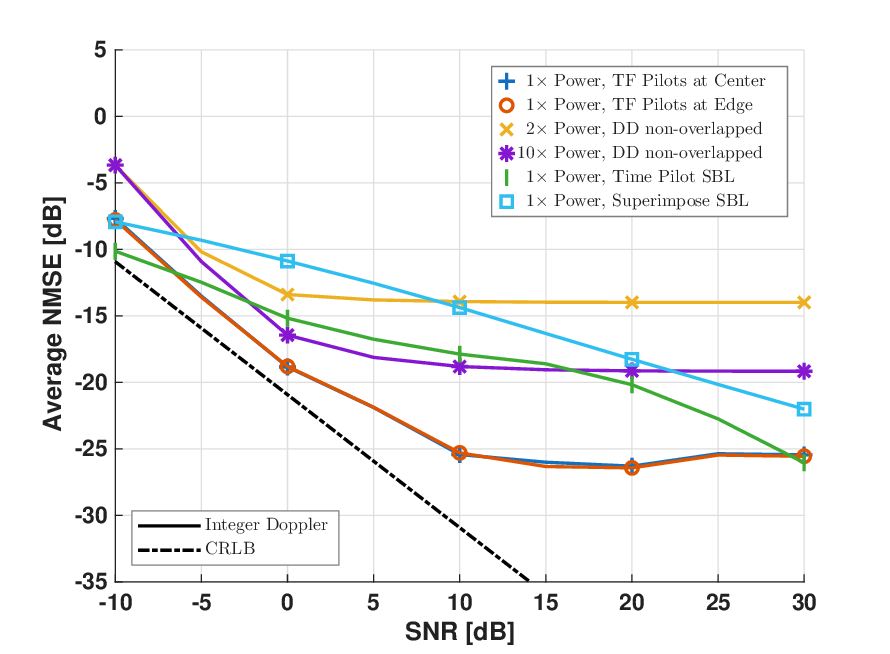}%
        \label{fig:nmse_baseline_integer}
    }%
    \subfloat[]{%
        \includegraphics[width=0.5\textwidth]
        {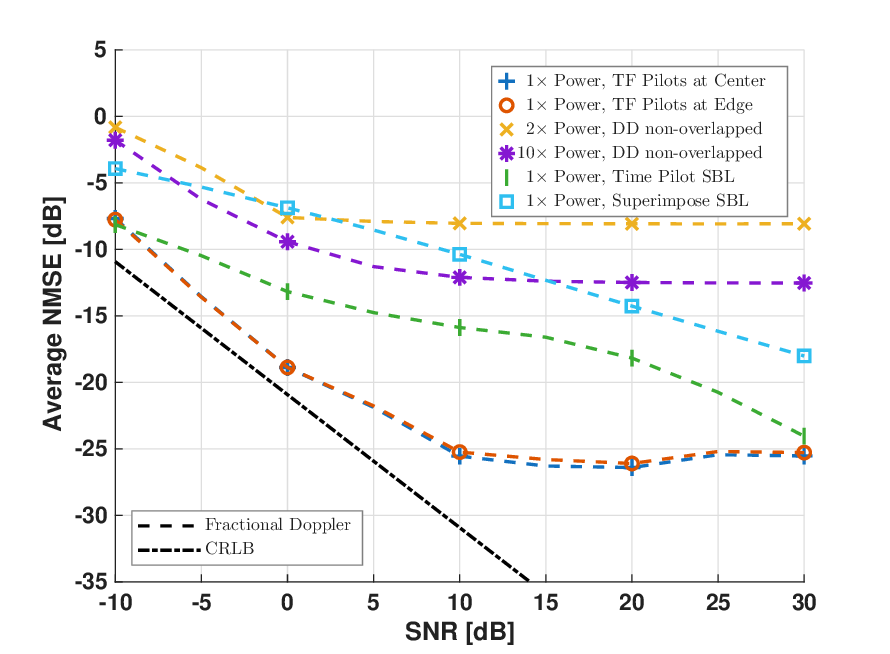}%
        \label{fig:nmse_baseline_fractional}
    }
    \caption{NMSE performance under (a) integer Doppler and (b) fractional Doppler channels.}
    \label{fig:nmse_baseline}
\end{figure*}

To investigate the NMSE performance of the proposed method, we perform the $500$-trial 
Monte Carlo simulations while varying the channel Doppler condition and SNR.
In each trial, the set of scatterer parameters is generated as described above.
In the integer Doppler case, the scatterers' Doppler indices are rounded to the nearest integer. 
The average NMSE performance is shown in \cref{fig:nmse_baseline}. 
For comparison, we evaluate threshold-based detection~\cite{raviteja2019EmbeddedPilotAided} \textcolor{black}{(noted as ``DD non-overlapped'' in the figure)}, setting its pilot power to twice and ten times the average symbol power, respectively. 
\textcolor{black}{Additionally, we evaluate the SBL estimator used in the superimposed-pilot scheme~\cite{mehrotra2023dataaidedCSI} (``Superimpose SBL'') and the time-domain pilot scheme~\cite{mehrotra2024onlinebayesian} (``Time Pilots SBL''), with both using $2048$ pilots ($1/32$ of the OTFS grid), and an oversampling factor of $4$ for fractional-Doppler estimation.}

We observe that the proposed method exhibits superior NMSE results across all SNR regimes, achieving a significantly lower NMSE than comparison methods. As expected, its performance is  
consistent
regardless of whether the Doppler shift is integer or fractional, whereas the methods in~\cite{raviteja2019EmbeddedPilotAided, mehrotra2023dataaidedCSI, mehrotra2024onlinebayesian} degrade under fractional Doppler conditions.
\textcolor{black}{
This is because the proposed method is based on \cref{thm:Y}, which provides a unified TF-domain expression for channel parameter estimation under both integer and fractional Doppler. In contrast, \cite{raviteja2019EmbeddedPilotAided, mehrotra2023dataaidedCSI, mehrotra2024onlinebayesian} estimate the DD-domain channel support and therefore suffer from fractional-Doppler-induced power leakage that is absent in the integer-Doppler case.}


\begin{figure}
    \centering
    \includegraphics[width=\linewidth]{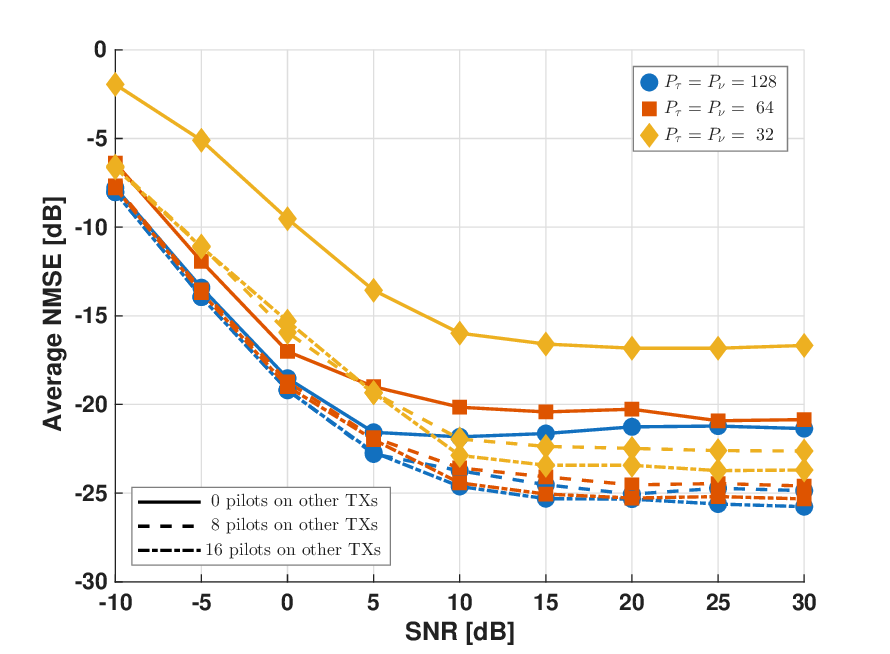}
    \caption{NMSE performance of different numbers of pilots.}
    \label{fig:nmse_npilots}
\end{figure}

Different pilot placements were also investigated, including configurations with pilots at the edges of the TF frame. Although \cref{fig:nmse_baseline} shows nearly identical NMSE performance for center- and edge-placed pilots, in practical implementations with non-rectangular pulse shaping and non-ideal filtering, edge-placed pilots may be more susceptible to leakage effects and more dependent on the CP configuration.

\begin{figure}
    \centering
    \includegraphics[width=\linewidth]{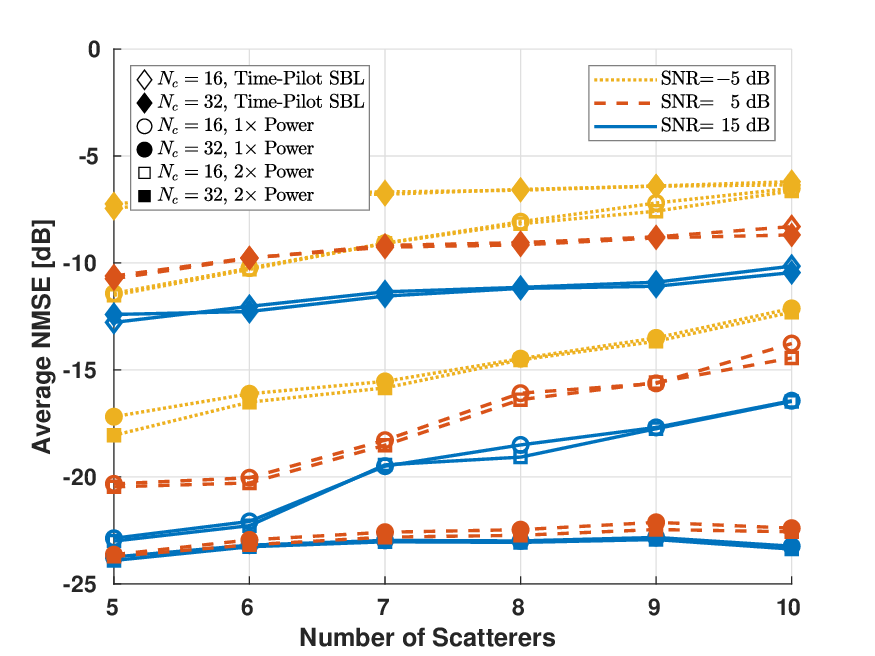}
    \caption{NMSE performance for different pilot powers and numbers of scatterers.}
    \label{fig:nmse_ntargets}
\end{figure}

As a benchmark, the Cram\'er--Rao lower bound (CRLB) for OTFS channel estimation~\cite{pan2021CramerRaoLow} with a single cyclic prefix and a $4\times 16$ MIMO system is calculated and plotted in \cref{fig:nmse_baseline}. 
Although a gap exists between the empirical NMSE and the CRLB for both systems, the proposed method approaches the CRLB more closely at low SNR. Note that the CRLB is calculated using the average pilot symbol power, which is consistent with our proposed system; this bound would naturally be lower if higher pilot powers were assumed for embedded-pilot methods. 
\textcolor{black}{The performance curves of the benchmark schemes~\cite{mehrotra2023dataaidedCSI, mehrotra2024onlinebayesian} do not follow the CRLB because the latter is derived for the proposed single-CP OTFS signal model, whereas these schemes employ a CP per subsymbol. Their per-subsymbol CP design also incurs greater overhead than the proposed design. Finally, compared to the proposed method, the computational complexity of the superimposed-pilot (SBL with SIC)~\cite{mehrotra2023dataaidedCSI} is $140$ times higher, while that of the time-domain pilot (SBL)~\cite{mehrotra2024onlinebayesian} is $20$ times higher.}


\subsubsection{Effect of Number of Pilots, Pilot Power, and Number of Scatterers}

\color{black}
To further investigate the impact of pilot overhead on estimation accuracy, we conducted additional Monte Carlo simulations by varying $P_{\tau}$, $P_{\nu}$, and $P_{\mathrm{random}}$. 
The NMSE performance, illustrated in \cref{fig:nmse_npilots}, reveals that for $P_{\mathrm{random}} = 0$, the NMSE saturates at a performance floor of $-22$ dB, regardless of the values of $P_{\tau}$ and $P_{\nu}$. 
This floor stems from the lack of pilots on auxiliary antennas, which prevents the algorithm from fully exploiting the spatial diversity gain offered by the multi-transmitter setup.
Notably, even this saturated performance exceeds that of the non-overlapped DD pilot method, which remains capped at $-20$ dB despite using ten times the pilot power (as shown in \cref{fig:nmse_baseline}).
\textcolor{black}{As $P_{\mathrm{random}}$ increases, the algorithm begins to effectively leverage $P_{\tau}$ and $P_{\nu}$ to reduce the NMSE by enhancing the resolvability of coarse delay/Doppler paths.} 
The results in \cref{fig:nmse_npilots} show that a minimum of $8$ pilots per auxiliary antenna (\eg, for TX$_3$ and TX$_4$, totaling $P_{\rm random}=16$) is necessary to break the performance floor. 
Beyond this threshold, further increases in $P_{\mathrm{random}}$ yield only marginal improvements. 
The optimal pilot counts and placement remain subjects for future study.

\textcolor{black}{We also performed Monte Carlo simulations with $500$ trials per setup to evaluate the effects of pilot power and the number of scatterers on the proposed method.} 
The results for the $4\times16$ and $4\times32$ systems are shown in \cref{fig:nmse_ntargets}. 
The proposed method remains robust as channel complexity increases; specifically, the system experiences a modest increase in NMSE of approximately $7$ dB when $J$ is doubled from $5$ to $10$.
Notably, for the $\textcolor{black}{N_c} = 32$ configuration, the NMSE remains nearly constant in the high-SNR regime. 
\textcolor{black}{
A larger \(\textcolor{black}{N_c}\) enhances coarse-stage AoA resolvability and, at the same time, improves SSR-based CSI recovery by increasing the number of dictionary rows from \(N_p\) to \(N_p\textcolor{black}{N_c}\), thereby providing a richer set of observations.}
\textcolor{black}{For comparison, we also evaluate the time-domain pilot scheme~\cite{mehrotra2024onlinebayesian}, which is less sensitive to the scatterer count owing to DD-support recovery, but does not scale well with $N_c$ due to pilot pollution; the superimposed-pilot scheme~\cite{mehrotra2023dataaidedCSI} is omitted from this test owing to its prohibitive computational complexity.}
Finally, we observe that setting the pilot power to the average transmission power yields only a marginal increase in NMSE compared to doubling it. 
\textcolor{black}{
This behavior is expected: for example, for the channel parameters listed in \cref{tab:channel_parameters} at an SNR of $15$~dB, the SINR increases from $13.37$~dB when the pilot power equals the average transmit power to $16.37$~dB when the pilot power is doubled.}
This suggests a key advantage: high performance can be maintained with a uniform power profile, simplifying hardware and power amplifier requirements.

\begin{figure}
    \centering
    \includegraphics[width=0.98\columnwidth]{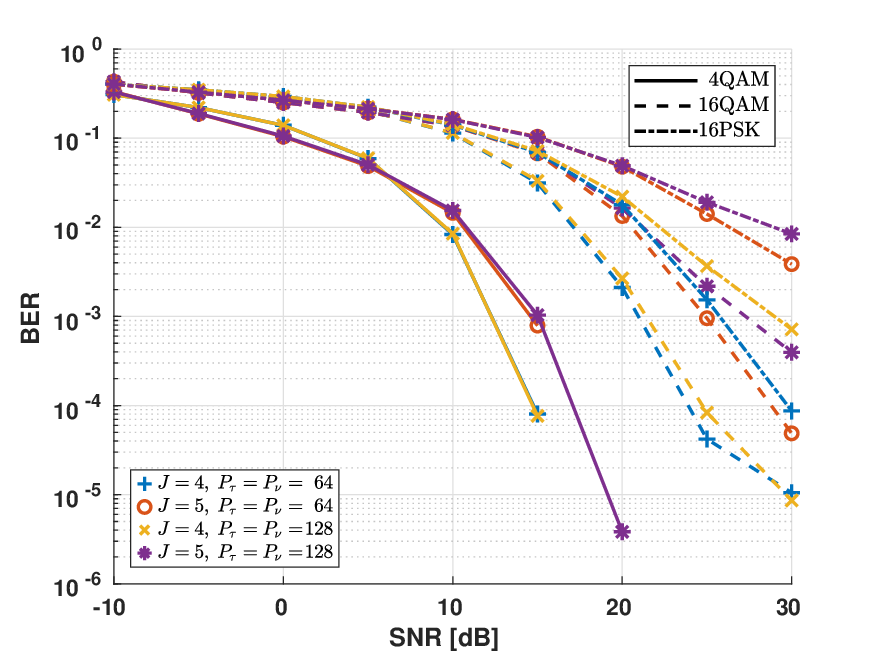}
    \caption{BER performance for various modulation, number of scatterers/pilots.}
    \label{fig:BER}
\end{figure}

\subsubsection{BER Performance}

We evaluate bit-error-rate (BER) performance under various modulation schemes, scatterer densities, and pilot configurations across a range of SNR levels.
As illustrated in \cref{fig:BER}, QPSK consistently outperforms 16QAM and 16PSK, as anticipated. 
A slight BER degradation is observed as the number of scatterers increases, consistent with the marginal increase in channel estimation NMSE in more complex scattering environments.  
One way to accommodate more scatterers is to increase $\textcolor{black}{N_c}$ (see \cref{fig:nmse_ntargets}).
Increasing the number of pilots yields diminishing NMSE gains and introduces additional residual interference (see \cref{eq:received_data}), leading to a slight increase in BER; however, \cref{fig:BER} shows that the resulting performance penalty remains small.

\subsubsection{Comparison With Overlapped DD-Pilot Schemes}

\textcolor{black}{
Here, we compare the proposed pilot design to the overlapped DD pilot designs with the 3D-OMP algorithm~\cite{shen2019ChannelEstimation} and with 3D-OMP using dedicated 2D-Pilots~\cite {liang2024TwoDimensionalDelayDoppler}, both of which consider downlink transmission from a multi-antenna base station (BS) to multiple single-antenna user equipment (UE) devices.}
The channel model of~\cite{shen2019ChannelEstimation, liang2024TwoDimensionalDelayDoppler} follows the 3GPP macrocell standard~\cite{BibEntry2026Mar}, where each scatterer spreads energy across $20$ subpaths in the angular domain.
The received TF-domain signal of each UE on the private bin $[n_p,m_p]$ of the $p$-th antenna is:
\begin{align}
    \frac{Y'[n_p,m_p]}{X_{p}[n_p,m_p] } 
    &= \sum_{j=0}^{J-1} \underbrace{\sum_{j^r=1}^{20} e^{-i2\pi pg_t \frac{\sin\varphi_{j^r}}{\lambda}} \beta_{j^r}}_{\mathcal{A}_j(p)} H^j[n_p,m_p] 
    . 
    \label{3GPP}
\end{align}
Here, $\beta_{j^r}$ is the complex channel gain of the $j^r$-th subpath associated with the $j$-th scatterer, and $\mathcal{A}_j(p)$ is the composite gain of the $j$-th path at transmit antenna $p$.
It can be seen that the term $H^j[n_p,m_p]$ does not depend on the antenna $p$, while the coefficient $\mathcal{A}_j(p)$ does depend on the antenna $p$. 
We can adopt the algorithms described in \cref{sec:coarse_estimates,sec:SSR} to estimate the Doppler and delay. The coefficients $\mathcal{A}_j(p)$ are then determined by solving the LS problem in \cref{3GPP}, requiring a minimum of $J$ pilots for each TX.

We first apply the approach described in \cref{doppler-estimation,delay-estimation} of \cref{sec:coarse_estimates} to obtain coarse estimates $\hat{\nu}_j$ and $\hat{\tau}_j$ based on the pilot arms of antenna $p$.
Then, by stacking the $Y'[n_p,m_p]/X_p[n_p,m_p]$ ratios from private bins into a vector $\mathbf{r}_p'$, the system model for the $p$-th antenna is formulated as $\mathbf{r}_p'=\mathbf{\Phi}_p'\bm{\beta}_p'$, where the dictionary $\mathbf{\Phi}_p'$ is constructed using discretized delay and Doppler parameters around coarse estimates. 
Solving this SSR via OMP with binary refinement yields refined estimates. 
Finally, for each antenna $p$, an LS solution is formulated using at least $J$ pilots along with the refined dictionary to estimate $\mathcal{A}_j(p)$ for each path $j$, thereby circumventing the estimation of individual $\beta_{j^r}$ values.

Applying the SFFT to \cref{3GPP} yields the following channel matrix for NMSE evaluation.
\begin{multline}
    \mathbf{h}_{\text{MIMO}}'
    = \sum_{j=0}^{J-1}\parens*{\sum_{j^r=1}^{20}\beta_{j^r} \mathbf{a}_{t}\hmt(\varphi_{j^r})} \otimes \\
    (\mathbf{F}_{\text{N}}\otimes\mathbf{I}_{\text{M}})
    (\Pi^{l_j}\Delta^{k_j+\kappa_j})
    (\mathbf{F}_{\text{N}}\hmt\otimes\mathbf{I}_{\text{M}}).
\end{multline}
The first term in parentheses corresponds to a vector of the composite path gains, $[\mathcal{A}_j(0), \dots, \mathcal{A}_j(N_t-1)]\trn$, capturing the combined effect of the 20 subpaths at each antenna.

\begin{figure}
    \centering
    \includegraphics[width=\linewidth]{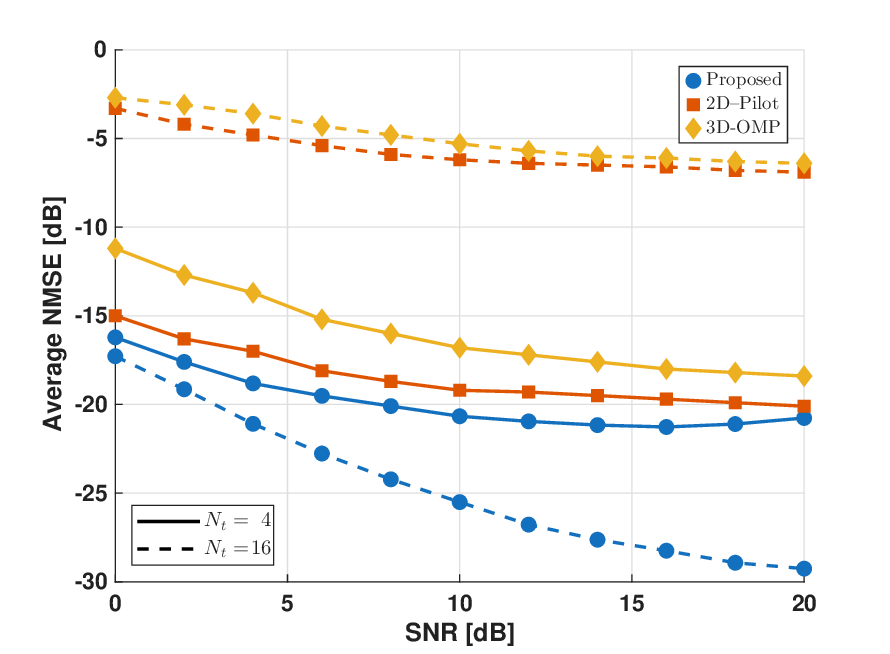}
    \caption{NMSE performance of the TF-pilot and overlapped DD-pilot schemes.}
    \label{fig:compare}
\end{figure}

Following the system setup ($N_t\in\{4,16\}$ and $M\times N=512\times 140$) and pilot setup ($121$ pilots) in~\cite{liang2024TwoDimensionalDelayDoppler}, we performed a $500$-trial Monte Carlo simulation. 
As shown in \cref{fig:compare}, at $N_t=4$, the proposed system only slightly outperforms both benchmarks, as the pilot pollution from overlapping structures is not severe when $N_t$ is small.
However, at $N_t=16$, both benchmarks suffer significant degradation from pilot pollution. 
The proposed orthogonal approach remains robust to pilot pollution and leverages the expanded pilot set at $N_t=16$ to substantially improve estimation precision.

The proposed algorithm also has substantially lower computational complexity.
The OMP algorithm's complexity is dominated by the correlation step, which scales with the dictionary matrix's dimensionality. 
For the method in~\cite{shen2019ChannelEstimation}, which performs $J$ OMP iterations, the complexity is $\mathcal{O}(J N_p N_t MN)$. 
The approach in \cite{liang2024TwoDimensionalDelayDoppler} utilizes the same estimator but includes a matched-filtering stage for its dedicated 2D pilots, resulting in a total complexity of $\mathcal{O}((L_{\tau}L_{\nu})^2 + J N_p N_t MN)$, where $L_{\tau} \times L_{\nu}$ is the 2D pilot grid size used in~\cite{liang2024TwoDimensionalDelayDoppler}.
The proposed method has a complexity of $\mathcal{O}(P_{\tau}\log P_{\tau}+P_{\nu}\log P_{\nu}+J N_{\rm iter} N_p N_tG_{\tau}G_{\nu})$, where the first two terms are the complexities of the fast Fourier transforms (FFTs) used in the two DFT-based coarse-estimation steps, and $N_{\rm iter}$ is the number of binary refinement iterations. 
Since $G_{\tau}=11 \ll M=512$ and $G_{\nu} =21 \ll N= 140$, the overall complexity is much lower than those of~\cite{shen2019ChannelEstimation, liang2024TwoDimensionalDelayDoppler}.

\color{black}

\section{Conclusion}\label{sec:conclusion}
\color{black}
We have shown that the TF domain, rather than the DD domain, is the natural framework for scalable, low-overhead MIMO-OTFS channel estimation. \cref{thm:Y}  establishes that under rectangular pulse shaping and matched filtering, each propagation path's TF-domain response reduces to a pure complex exponential across the subsymbol index, scaled by a subsymbol-independent coefficient with identical structure under integer and fractional Doppler, while the residual interference is exactly characterized.
This Doppler invariance, absent from the DD representation (\cref{thm:y}), enables low-complexity coarse parameter estimation directly in TF domain.

Building on this result, we proposed a TF-domain pilot design using private TF bins with TF and DD guard bins, whose placement is guaranteed invertible via an explicit rank condition. The design decouples pilot overhead from the number of transmit antennas and admits a closed-form SINR characterization of the resulting interference. The associated coarse-to-fine estimation framework, consisting of DFT-based coarse estimation, virtual-array construction, and a dimensionality-reduced sparse recovery procedure, achieves high-resolution angle, delay, Doppler, and gain estimates at a fraction of the complexity of direct 4D sparse recovery.

Simulations confirmed consistent NMSE and BER performance under both integer and fractional Doppler, close approach to the CRLB, robustness to pilot pollution as the number of transmit antennas grows, and substantially lower pilot overhead and complexity than existing baselines. Future work will address optimal pilot placement beyond the arm-based design, and extend the framework to more general propagation environments, including unknown transmitter–receiver geometry and synchronization errors.

\color{black}

\appendices
\crefalias{section}{appendix} 
\section{}\label{sec:Appendix_Y}

Following the analysis of~\cite{raviteja2018InterferenceCancellation}, with rectangular pulses, each received subsymbol is only affected by the current and the previous subsymbol, \ie,
\begin{align}
&Y[n,m]
= \sum_{\substack{m'=0}}^{M-1} X[n,m'] H_{n,m}[n,m'] \label{eq:sigici}\\
&+ \sum_{m'=0}^{M-1} X[n-1,m'] H_{n,m}[n-1,m']+W[n,m] \label{eq:sigisi} \\
&= Y^{\rm ici}[n,m]+Y^{\rm isi}[n,m]+W[n,m], \label{eq:sigiciisi}
\end{align}
where $Y^{\rm ici}[n,m]$ contains the signal and inter-carrier interference (ICI),
and $Y^{\rm isi}[n,m]$ contains the intersubsymbol interference (ISI) from the previous subsymbol. 


On substituting~\cite[Eq. (22)]{raviteja2018InterferenceCancellation} into $X[n,m]H_{n,m}[n,m]$, and letting $p=l-l_j$, the received TF signal on bin $[n,m]$ is
\begin{align}
    &X[n,m]H_{n,m}[n,m] \\
    &= X[n,m] \sum_{j=0}^{J-1} \beta_j e^{-i2\pi\parens*{\nu_j+m\Delta f}\tau_j}e^{i2\pi\nu_j n\Delta t} \\
    &\quad \times \frac{1}{M} \sum_{p=0}^{M-1-l_j} e^{i2\pi\nu_j\parens*{\frac{p}{M\Delta f}+\tau_j}} \label{eq:40}\\
    &= X[n,m] \sum_{j=0}^{J-1} \beta_j e^{-i2\pi\frac{(k_j+\kappa_j)l_j}{NM}} e^{i2\pi\bracks*{\frac{(k_j +\kappa_j)n }{N} - \frac{ml_j}{M}}} \xi_j, \label{eq:53}
\end{align}
where (from \cref{eq:40})
\begin{align*}
    \xi_j
    &= \frac{1}{M}\sum_{l=l_j}^{M-1}e^{i2\pi\frac{(k_j+\kappa_j)l}{NM}} 
\end{align*}
leading to \cref{eq:xi_rest}.
Similarly, we can derive the ICI (using~\cite[Eq. (22)]{raviteja2018InterferenceCancellation} and letting $p=l-l_j$),
and the ISI (using~\cite[Eq. (23)]{raviteja2018InterferenceCancellation} and letting $p=l+(M-l_j)$).
The sum of the two terms is the $I[n,m]$ term of \cref{eq:TFinterference}.


\cref{thm:Y} can also be derived from \cref{thm:y} using the technique developed in~\cite{wang2025ISACMIMO}.
Specifically, 
%
extending the summation in \cite[Eq. (32)]{wang2025ISACMIMO} to $q\in[k_j-(N-1),k_j]$ reduces it to $\xi_j e^{i 2\pi \frac{\kappa_j n}{N}}$, yielding \cref{eq:53} and completing the proof.
\section{}\label{sec:appendix_I_SISO}

\color{black}

The $I[n,m]$ of \cref{eq:TFinterference} can be decompose as follows,
\begin{align}
&I[n,m]=I_{0}[n,m]+I_{-1}[n,m],  \\
&I_{0}[n,m]=\sum_{\substack{m'=0\\ m'\neq m}}^{M-1} g_0[n,m,m']X[n,m'],  \\
&I_{-1}[n,m]=\sum_{m'=0}^{M-1}g_{-1}[n,m,m']X[n-1,m'],  \\
&g_0[n,m,m']= \sum_{j=0}^{J-1}
\beta_j e^{-i2\pi\frac{(k_j+\kappa_j)l_j}{NM}}
e^{i2\pi\frac{(k_j+\kappa_j)n}{N}}  \\
&\quad\times\frac{1}{M} e^{-i2\pi\frac{m'l_j}{M}} \sum_{l=l_j}^{M-1}
e^{i2\pi\bracks*{\frac{(k_j+\kappa_j)l}{NM}-\frac{(m-m')l}{M}}},  \\
&g_{-1}[n,m,m']= \sum_{j=0}^{J-1}
\beta_j e^{-i2\pi\frac{(k_j+\kappa_j)l_j}{NM}}
e^{i2\pi\frac{(k_j+\kappa_j)n}{N}}  \\
&\quad\times\frac{1}{M} e^{-i2\pi\frac{m'l_j}{M}} \sum_{l=0}^{l_j-1}
e^{i2\pi\bracks*{\frac{(k_j+\kappa_j)l}{NM}-\frac{(m-m')l}{M}}}. 
\end{align}


For notational simplicity, we temporarily omit the index $[n,m]$. 
Since the TF samples are mutually uncorrelated, 
the cross terms vanish, and the interference power becomes 
\begin{align}
\E{\abs{I}^2}&=
 E\left[|I_0|^2\right]+E\left[|I_{-1}|^2\right] \label{eq:interference_twopart}
\end{align}

Since $\E{\beta_j\beta_{j'}^*}=\sigma_{\beta}^2\delta[j-j']$, we obtain
\begin{align}
&\E{\abs{I_{0}}^2} = \E{\E{\abs{I_{0}}^2\mid \bm{\beta}}}, \\
&= \mathbb{E}\Biggl[\sum_{m'\neq m}\sum_{m''\neq m}g_0[m']g_0^*[m''] 
\E{X[n,m']X^*[n,m'']\mid \bm{\beta}}\Biggr] \notag \\
&=\E{\sum_{m'\neq m}\abs{g_0[m']}^2} \label{eq:66} \\
&=\frac{\sigma_{\beta}^2}{M^2}\sum_{j=0}^{J-1}\sum_{m'\neq m} \abs*{\sum_{l=l_j}^{M-1} 
e^{i2\pi\bracks*{\frac{(k_j+\kappa_j)l}{NM}-\frac{(m-m')l}{M}}}}^2. \label{eq:070} \\
%
&\E{\abs{I_{-1}}^2} 
=\E{\sum_{m'=0}^{M-1}\abs{g_{-1}[m']}^2} \label{eq:67} \\
&=\frac{\sigma_{\beta}^2}{M^2}\sum_{j=0}^{J-1}\sum_{m'=0}^{M-1} \abs*{\sum_{l=0}^{l_j-1} 
e^{i2\pi\bracks*{\frac{(k_j+\kappa_j)l}{NM}-\frac{(m-m')l}{M}}}}^2. \label{eq:72}
\end{align}

To simplify these expressions, let $q=[m-m']_M$ and
\begin{align}
F_{0,j}[l]=
\begin{cases}
    e^{i2\pi \frac{(k_j+\kappa_j)l}{NM} }, & l=l_j,l_j+1,\ldots,M-1, \\
    0, & \text{otherwise.}
\end{cases}
\end{align}
Temporarily including the term $m'=m$, \cref{eq:070} gives
\begin{align}
&\sum_{m'=0}^{M-1}\abs*{\sum_{l=l_j}^{M-1} 
e^{i2\pi\bracks*{\frac{(k_j+\kappa_j)l}{NM}-\frac{(m-m')l}{M}}}}^2 \\
&= \sum_{q=0}^{M-1}\abs*{\sum_{l=0}^{M-1} F_{0,j}[l]e^{-i2\pi q \frac{l}{M}} }^2
= M\sum_{l=0}^{M-1}\abs{F_{0,j}[l]}^2 \label{eq:Paeseval_2} \\
&= M(M-l_j). \label{eq:Paeseval_3}
\end{align}
Here, \cref{eq:Paeseval_2} follows Parseval's theorem, while \cref{eq:Paeseval_3} follows because $\abs{F_{0,j}[l]}=1$ for exactly $M-l_j$ samples and is zero elsewhere. Similarly,
\begin{align}
\sum_{m'=0}^{M-1} \abs*{\sum_{l=0}^{l_j-1} 
e^{i2\pi\bracks*{\frac{(k_j+\kappa_j)l}{NM}-\frac{(m-m')l}{M}}}}^2 = Ml_j. \label{eq:Paeseval_4}
\end{align}
The actual interference sum excludes the term $m'=m$, \ie, $q=0$, whose contribution is
\begin{align}
    \E{\abs*{g_0[n,m,m]}^2}
    &\!=\! \E{\abs*{\sum_{j=0}^{J-1}\beta_jH^j[n,m]\xi_j}^2}
     \!=\! \sigma_{\beta}^2\sum_{j=0}^{J-1}\abs*{\xi_j}^2. \label{eq:excludedterm}
\end{align}

Substituting \cref{eq:Paeseval_3,eq:Paeseval_4,eq:excludedterm} into \cref{eq:070,eq:72} and then using \cref{eq:interference_twopart} yields
\begin{align}
\E{II^*} 
&\!=\! \sigma_{\beta}^2\sum_{j=0}^{J-1}\bracks*{\frac{M(M-l_j)+Ml_j}{M^2} -\abs{\xi_j}^2} \\
&\!=\! \sigma_{\beta}^2\sum_{j=0}^{J-1}\bracks*{1-\abs{\xi_j}^2}. \label{eq:includedterm}
\end{align}

\color{black}
\section{}\label{sec:appendix_y}

Applying the SFFT to $Y[n,m]$ (\cref{eq:sigiciisi}) yields $y[k,l]$
which can be expressed as $y[k,l] = y^{\mathrm{ici}}[k,l] + y^{\mathrm{isi}}[k,l]+w[k,l]$.
\begin{table*}
    \centering
    \begin{minipage}{\textwidth}
    \begin{align}
    y^{\mathrm{ici}}[k,l]
    &= \frac{1}{N}\sum_{j=0}^{J-1} \beta_j
    \Biggl[\sum_{l'=0}^{M-1}\sum_{t''=0}^{M-1-l_j}
    e^{i2\pi\frac{t''}{M}\left(\frac{k_j+\kappa_j}{N}\right)}
    \delta\bigl([t''+l_j-l]_M\bigr)
    \delta\bigl([t''-l']_M\bigr) 
    \sum_{\substack{q=k_j-\\ (N-1)}}^{k_j}\gamma_j(q)x\bigl[[k-(k_j-q)]_N, t''\bigr]
    \Biggr]. \label{eq:70} \\
    y^{\mathrm{ici}}[k,l] &=
    \begin{cases}
    \displaystyle
    \sum_{j=0}^{J-1}\sum_{\substack{q=k_j-\\ (N-1)}}^{k_j} \beta_j
    \left[\frac{1}{N}\gamma_j(q)\right]
    e^{i2\pi\left(\frac{l-l_j}{M}\right)\left(\frac{k_j+\kappa_j}{N}\right)}
    x\bigl[[k-(k_j-q)]_N,[l-l_j]_M\bigr],
    & l \geq l_j,\\
    0, & \text{otherwise}.
    \end{cases} \label{eq:yici}
    \end{align}
    \hrule
    \end{minipage}
\end{table*}

\begin{table*}
\centering
\begin{minipage}{\textwidth}
\begin{align}
y^{\mathrm{isi}}[k,l]
&= \frac{1}{N}\sum_{j=0}^{J-1} \beta_j
   \Biggl[\sum_{t''=M-l_j}^{M-1}
   e^{i2\pi\left(\frac{t''-M}{M}\right)\left(\frac{k_j+\kappa_j}{N}\right)}
   \delta\bigl([t''+l_j-l]_M\bigr) 
   \sum_{\substack{q=k_j-\\ (N-1)}}^{k_j}
   \bigl(\gamma_j(q)-\textcolor{black}{1}\bigr)
   \textcolor{black}{e^{-i2\pi\frac{[k-(k_j-q)]_N}{N}}}
   x\bigl[[k-(k_j-q)]_N,t''\bigr]
   \Biggr]. \label{eq:76}\\
y^{\mathrm{isi}}[k,l] 
&=
\begin{cases}
\displaystyle
\sum_{j=0}^{J-1}\sum_{\substack{q=k_j-\\ (N-1)}}^{k_j} \beta_j
\left[\frac{1}{N}\bigl(\gamma_j(q)-1\bigr)\right]
e^{-i2\pi\frac{[k-(k_j-q)]_N}{N}}
e^{i2\pi\left(\frac{l-l_j}{M}\right)\left(\frac{k_j+\kappa_j}{N}\right)} 
x\bigl[[k-(k_j-q)]_N, [l-l_j]_M\bigr],
& l < l_j,\\
0, & \text{otherwise}.
\end{cases} \label{eq:yisi}
\end{align}
\hrule    
\end{minipage}
\end{table*}
For $y^{\mathrm{ici}}[k,l]$, we follow the proof technique outlined in~\cite[Eqs. (53)–(57)]{raviteja2018InterferenceCancellation}, but depart from the approximation in~\cite[Eq. (58)]{raviteja2018InterferenceCancellation} to establish an exact equality. 
Specifically, we let $k'=[k-(k_j-q)]_N$ for $q\in[k_j-(N-1),k_j]$ and define $\gamma_j(q)\stackrel{\triangle}{=}\sum_{n=0}^{N-1}e^{-i2\pi(-q-\kappa_j)\frac{n}{N}}$. 
This substitution yields \cref{eq:70}.
Observing that $y^{\mathrm{ici}}[k,l]$ is non-zero only if $l \geq l_j$ and $t'' = [l - l_j]_M$, we arrive at the final expression in \cref{eq:yici}.

Similarly, for $y^{\mathrm{isi}}[k,l]$, we follow the proof technique in~\cite[Eqs. (60)-(61)]{raviteja2018InterferenceCancellation}, but depart from the approximation in~\cite[Eqs. (62)-(63)]{raviteja2018InterferenceCancellation} to establish an exact equality with the same change of variables. The substitution yields \cref{eq:76}, which is non-zero only if $l < l_j$ and $t'' = [l - l_j + M]_M$ and results in \cref{eq:yisi}.
Combining \cref{eq:yici,eq:yisi}, we obtain \cref{thm:y}. 
\section{}\label{sec:appendix_I_short}

\color{black}

Omitting noise, the TF signal received at antenna $n_c$ on the private pilot
$[n_p,m_p]$ of antenna $p$ is
\begin{align}
    Y_{n_c}[n_p,m_p] &= X_p[n_p,m_p]\sum_{j=0}^{J-1}
    e^{-i2\pi n_c g_c \frac{\sin\theta_j}{\lambda}}
    e^{-i2\pi p g_t \frac{\sin\varphi_j}{\lambda}} \nonumber\\
    &\quad\times \beta_j H_j[n_p,m_p]\xi_j
    + I_{n_c}[n_p,m_p]. \label{eq:D_received}
\end{align}

Comparing with \cref{eq:TFinterference}, the contribution $I_{n_t}[n_p,m_p]$ of transmit antenna $n_t$ to $I_{n_c}[n_p,m_p]=\sum_{n_t=0}^{N_t-1}I_{n_t}[n_p,m_p]$ takes exactly the form of $I[n,m]$ from \cref{sec:appendix_I_SISO}, evaluated at $[n_p,m_p]$, with every term of the summand additionally multiplied by the unit-modulus steering vector
factors $e^{-i2\pi n_cg_c\sin\theta_j/\lambda} e^{-i2\pi n_tg_t\sin\varphi_j/\lambda}$. 
Since these factors have unit modulus, they leave every magnitude-squared quantity in \cref{sec:appendix_I_SISO}'s derivation unchanged, so $E[|I_{n_t}|^2]=E[|I|^2]$ for every $n_t$, with $E[|I|^2]$ given by \cref{eq:includedterm}.

We temporarily omit the index $[n_p,m_p]$. 
Assuming samples transmitted from different antennas are uncorrelated, the cross-antenna terms vanish, and the interference power becomes
\begin{equation}
    E\big[I_{n_c}I_{n_c}^*\big]
    = \sum_{n_t=0}^{N_t-1} E\big[|I_{n_t}|^2\big]
    = N_t\sigma_\beta^2\sum_{j=0}^{J-1}\big(1-|\xi_j|^2\big),
    \label{eq:D_interference}
\end{equation}
where the last equality follows from \cref{eq:includedterm}.

The desired signal power, {shown in the numerator of \cref{eq:SINR_simple}},
follows from \cref{eq:excludedterm} after replacing $\sigma_\beta^2$ with
$P_{\rm Pilot}\sigma_\beta^2$ to account for the pilot power, and is
independent of $[n_p,m_p]$, $p$, and $n_c$ since the steering-vector
and $H_j[n_p,m_p]$ factors have unit modulus.
The SINR, defined as the ratio of the desired signal power to the interference-plus-noise power (\cref{eq:D_interference}) plus $\sigma_w^2$, is \cref{eq:SINR_simple}.

\color{black}

\bibliographystyle{IEEEtran}
\bibliography{bib,ref}
\end{document}